\documentclass{aa}

\usepackage{graphicx}
\usepackage[varg]{txfonts}
\usepackage{longtable}
\usepackage{ragged2e}
\usepackage{pdflscape}

\begin{document}

\title{\textit{Gaia} DR2 study of Herbig Ae/Be stars}

\author{M. Vioque\inst{1,2}\thanks{\email{pymvdl@leeds.ac.uk}} \and R. D. Oudmaijer\inst{1} \and D. Baines\inst{3} \and I. Mendigut\'ia\inst{4} \and R. P\'erez-Mart\'inez\inst{2}}
\institute{School of Physics and Astronomy, University of Leeds, Leeds LS2 9JT, UK.
         \and
             Ingenier\'ia de Sistemas para la Defensa de Espa\~na (Isdefe), XMM/Newton Science Operations Centre, ESA-ESAC Campus, PO Box 78, 28691 Villanueva de la Ca\~nada, Madrid, Spain.
         \and
             Quasar Science Resources for ESA-ESAC, ESAC Science Data Center,  PO Box 78, 28691 Villanueva de la Ca\~nada, Madrid, Spain.
         \and
             Centro de Astrobiolog\'ia (CSIC-INTA), Departamento de Astrof\'isica, ESA-ESAC Campus, PO Box 78, 28691 Villanueva de la Ca\~nada, Madrid, Spain.                               
             }

\date{Accepted for publication in Astronomy \& Astrophysics.}
\abstract
{}
{We use {Gaia Data Release 2 (DR2)} to place {252} Herbig Ae/Be
stars in the HR diagram and investigate their characteristics and
properties.}
{For all known Herbig Ae/Be stars with parallaxes in {Gaia DR2}, we collected their atmospheric parameters and photometric and extinction values from the literature.  To these data we added near- and mid-infrared
  photometry, collected H$\alpha$ emission line properties such as
  equivalent widths and line profiles, and their binarity status. In
  addition, we developed a photometric variability indicator from {Gaia's
  DR2} information.}
{We provide masses, ages, luminosities, {distances}, {photometric variabilities} and infrared
  excesses homogeneously derived for the most complete sample of
  Herbig Ae/Be stars to date.  We find that high mass stars have a
  much smaller infrared excess and have much lower optical
  variabilities compared to lower mass stars, with the break at around
  7M$_{\odot}$.  H$\alpha$ emission is generally correlated with
  infrared excess, with the correlation being stronger for infrared
  emission at wavelengths tracing the hot dust closest to the
  star. The variability indicator as developed by us shows that
  {$\sim$25\%} of all Herbig Ae/Be stars are strongly variable. We
  observe that the strongly variable objects display doubly peaked
  H$\alpha$ line profiles, indicating an edge-on disk.}
{ The fraction of strongly variable Herbig Ae
  stars is close to that found for {A-type} UX Ori stars. It had been
  suggested that this variability is in most cases due to asymmetric
  dusty disk structures seen edge-on. The observation here is in
  strong support of this hypothesis. Finally, the difference in dust
  properties occurs at 7M$_{\odot}$, while various
  properties traced at UV/optical wavelengths differ at a lower mass,
  3M$_{\odot}$. The latter has been linked to different accretion
  mechanisms at work whereas the differing infrared properties and {photometric variabilities} are related
  to different or differently acting (dust-)disk dispersal mechanisms. 
}


\maketitle

\section{Introduction} \label{sec:intro}

Herbig Ae/Be stars (HAeBes) are Pre-Main Sequence stars (PMS) of
intermediate mass, spanning the range between low mass T-Tauri stars
and the embedded Massive Young Stellar Objects (MYSOs). They are
optically bright so they are much easier to observe and to study than
MYSOs and it is expected that within the mass range of HAeBes a change
in accretion mechanism from the magnetically controlled accretion
acting for T-Tauri stars (see \citealp{Bouvier}) to an, as yet,
unknown mechanism for high mass stars occurs. Indeed, there is
evidence that the magnetically driven accretion model is valid for
Herbig Ae stars but not for several Herbig Be stars
(\citealp{Fairlamb}; \citealp{Ababakr}; \citealp{Oudmaijer2};
\citealp{Grady}; \citealp{Scholler}). Moreover, there are multiple
evidences that Herbig Ae and T-Tauri stars behave more similarly than
Herbig Be stars, and Herbig Ae and Herbig Be stars have different
observational properties. Examples of this are the different outer gas
dispersal rates (higher for Herbig Be stars, \citealp{Fuente}), the
higher incidence of clustering scenarios for Herbig Be stars
(\citealp{Testi}), and the evidences of Herbig Be stars hosting denser
and larger inner gaseous disks (\citealp{Ilee}; \citealp{Monnier})
that may suggest a different accretion scenario with the disk reaching
directly into the star (\citealp{Kraus2}). Other spectro-photometric
(\citealp{Mendigutia4}; \citealp{Cauley} and \citealp{Patel}) and
spectro-polarimetric studies (\citealp{Vink2}) also point in the
direction that the accretion physics change within the Herbig Ae/Be
stars mass range. In addition, Herbig Be stars are more likely to be
found in binaries than Herbig Ae stars (\citealp{Baines}).

An important indicator of their PMS nature, {together with emission lines}, is the infrared (IR)
excess that also traces the Herbig Ae/Be forming environment. The IR
excess profile have been classified into two groups differentiated by
a flat or rising shape of the continuum (\citealp{Meeus}). This
difference has a geometric origin depending on the presence of flaring
outer disks and puffed-up inner disks (\citealp{Dullemonda}, 2004b, 2005),
 and the presence of gaps in the disk (\citealp{Maaskant}; \citealp{Honda}).
The IR excess of HAeBes is expected to be characteristic
and different from the IR excess of other similar objects like for
example, ordinary Be stars \citep{Finkenzeller}.

Herbig Ae/Be stars are known to present irregular photometric
variations, with a typical timescale from days to weeks
(\citealp{Eiroa}; \citealp{Oudmaijer5}) and of the order of one
magnitude in the optical. This variability is typically understood as
due to variable extinction, due to for example rotating circumstellar
disks, or as an effect of rotation on cold photospheric spots and also
pulsation due to the source crossing the instability strip in the HR
diagram (\citealp{Marconi}). An extreme case of large non-periodic
photometric and polarimetric variations is observed in UX Ori type
stars (UXORs) with amplitudes up to $2-3$ mag. Many of them are
catalogued as HAeBes and their extreme variability is explained by
eclipsing dust clouds in nearly edge-on sources and the scattering
radiation in the circumstellar environment (see \citealp{Grinin} and
references therein; \citealp{Natta} and \citealp{Natta2}).

Infrared photometric variability, related to disk structure
variations, is not always correlated with the optical variability
(\citealp{Eiroa}) which implies that different mechanisms regarding
both the disk structure and accretion underlie the final observed
variability. Spectroscopic variability is also present in Herbig Ae/Be
stars (\citealp{Mendigutia3}).

With the advent of the {second} data release of Gaia (DR2, \citealp{Gaia Collaboration2016}, \citealp{Gaia Collaboration2018}),
 providing parallaxes to over {1.3 billion} objects ({\citealp{Lindegren_new}}), 
 including the majority of known Herbig
Ae/Be stars, the time is right for a new study on the properties of
the class. {Gaia DR2} contains a five dimensional astrometric
solution ($\alpha$, $\delta$, $\mu_\alpha$, $\mu_\delta$ and parallax ($\varpi$)) 
up to $G\lesssim21$ (white G band,
described in {\citealp{Evans}}).
 {Almost all} of the known Herbig Ae/Be stars have parallaxes {in Gaia DR2},
 which allowed luminosities to be derived and {252} HAeBes to
be placed in the HR diagram, a {tenfold} increase on earlier studies
using Hipparcos data alone.



The paper is organized as follows: In Sect. \ref{sec:Data_acq}, we describe the data
acquisition of not only the parallaxes, but also optical and infrared
photometry, effective temperatures, extinction values, H$\alpha$ emission line information and binarity.
In Sect. \ref{sec:Der_cuantities} we derive the stellar luminosities and place
the objects in an Hertzsprung-Russell (HR) diagram, while we also present a method to derive
a statistical assessment of the objects' variability in Gaia's
database. In addition, we homogeneously derive masses and ages for all
the sources, together with near- and mid-infrared excesses.
In Sect. \ref{sec:Data_analysis} we carry out an analysis of the data and
present various correlations and interdependencies, which we discuss
in the context of intermediate mass star formation in Sect. \ref{sec:Discussion}. We
conclude in section Sect. \ref{sec:Conclusions}.

\section{Data acquisition}\label{sec:Data_acq}

\subsection{Construction of the sample}\label{sec:Construction}

We have gathered the majority of Herbig Ae/Be stars known and proposed to date from
different works ({272}, see \citealp{Chen} for a compilation of most of
them). \citet{Chen} based their sample mostly on the work of
\citet{Zhang} which in turn is based on the work of
\citet{The} and \citet{Vieira}. In addition, we included a few HAeBes
from {\citet{Alecian}; \citet{Baines}; \citealp{Carmona}; \citealp{Fairlamb}; 
\citet{Hernandez2}; \citet{Manoj} and \citet{Sartori}}
that are not present in the aforementioned papers.

Although Herbig Ae/Be stars have long been considered, by definition,
of type A or B, there should be some flexibility in this
constraint as the physical boundary between Herbig Ae stars and
intermediate mass T-Tauris is fairly unstudied. This is because
spectral types of T-Tauri stars are typically K-M with some G-type
objects while Herbig Ae/Be stars are, quite unsurprisingly, limited to A and
B spectral type. Hence, pre-Main Sequence stars of intermediate
spectral types have often been largely understudied. We therefore keep
objects with F-type classification in \citet{Chen} in the
sample. Similarly, no upper limit in mass was imposed, leaving the
separation between Massive Young Stellar Objects (MYSOs) and HAeBes to the
optical brightness of the sources\footnote{The MYSOs are typically
  infrared-bright and optically faint (\citealp{Lumsden}). However, a
  number of optically visible objects are known to have passed all
  selection criteria such as the early type objects \object{PDS 27} and \object{PDS 37}
  that are also classified as Herbig Be stars
  (\citealp{Ababakr3}).}.

Then, we crossmatched the sources with {Gaia DR2}. Detections were
considered to be matched with the catalogue when their coordinates
agreed to within {0.5} arcsecond. {If more than one match was found we took the closest one. 
If no match was found within 0.5 arcsecond, successive crossmatches with larger apertures were performed up to 2 arcsecond. In these latter cases
an individual inspection of the crossmatch was applied. Finally, a comparison between the Johnson V band magnitudes
and the Gaia filters was done for each source in order to discard possible incorrect matches.}
This provides us with parallaxes for {254} HAeBes.

 

{As \citet{Lindegren_new} point out, not all Gaia DR2
  parallaxes are of the same quality, and some values - despite their
  sometimes very small error bars - appear erroneous
  (\textit{e.g.} \citealt{Lindegren_new}). We included the following constraint
  in astrometric quality following the indications in Appendix C of
  \citet{Lindegren_new} and what was applied in \citet{HRdiagram}. This
  constraint will remove from the sample objects with spurious
  parallaxes:}

\begin{equation}\label{astrom}
u<1.2 \times {\rm max}(1,e^{-0.2(G-19.5)})
\end{equation}

{where $G$ is the Gaia G band and $u$ is the unit weight error,
  defined as the square root of the ratio of the astrometric$\_$chi2$\_$al and
  (astrometric$\_$n$\_$good$\_$obs$\_$al $-$ $5$) columns (\citealt{Lindegren_new}, their
  Equation C.2).  228 of our sources satisfy this condition.}

{Some objects are found to be very close to this
  condition, \object{PDS 144S}, \object{PV Cep}
  and \object{V892 Tau}, and as we will show later, they would appear significantly
  below the Main Sequence in the HR-diagram. Given that the Lindegren
  condition is presented as a guideline rather than a rule by the Gaia
  astrometry team, we decided to treat these three objects as
  if they satisfy Eq.~\ref{astrom} as well.}
  
{We will refer to the set of astrometrically well behaved sources as
the high quality sample and to  those that do not satisfy Eq.~\ref{astrom}
as the low quality sample. We will not be able to place 2
sources in the HR diagram because lacking of appropriate parameters (Sect. \ref{sec:Atmospheric}).
In addition, we will move 5 more sources to the low quality sample
in Sect. \ref{sec:HRdiagram} because of different reasons. 
Summarizing, there are 218 objects ($228-3-2-5$) in the final high quality sample and 34 in the low quality one.
Information about the objects in different samples is presented in separated tables
at the end of the paper.
The high quality sample will be the one taken into
account in further considerations unless otherwise specified.}  

Distances are not obtained by straightforwardly inverting the
parallax.  The conversion of one parameter to the other one is not
strictly trivial because of the non-linearity of the inverse function
(see for example \citealp{Bailer-Jones}). {In the case of Gaia
  DR2, \citet{Bailer-Jones_new} proposed distance values using a weak
  distance prior that follows a galactic model}. Their distances begin
to differ from the distances obtained through simple inversion for
sources with large errors, $\sigma_{\varpi}/\varpi\gtrsim0.5$.  Hence,
in our {initial (high and low quality)} sample only a small subset of {12}
Herbig Ae/Be stars will suffer substantially from this effect.
{Following the indications in \citet{Luri} on how to treat the
  Gaia parallaxes we decided to apply a simpler exponentially
  decreasing prior to estimate distances.}  {For completeness,
  we should note that the parallaxes provided by Gaia DR2 have a
  regional, not gaussian systematic error as large as $0.1$mas and a
  global zero point error of about $-0.029$mas that are not included
  in the gaussian random errors provided in the Gaia archive (see
  \citealp{Arenou} and \citealp{Lindegren_new}).  Hence, the
  uncertainty in the parallaxes is slightly underestimated.  The final
  errors in the high quality sample range from $0.016$ to $0.37$mas.}



Herbig Ae/Be stars have been historically confused with classical Be
stars, with which they share many characteristics (\citealp{Rivinius};
\citealp{Klement}; \citealp{Grundstrom}). Indeed, the nature of some
of the objects in our sample is still under debate. An interesting
example in this respect is \object{HD 76534}, a B2Ve object that
appears in listings of Be stars (eg. \citealp{Oudmaijer4}) and Herbig
Be stars alike (\citealp{Fairlamb}). The latest dedicated study puts
the object in the Herbig Be category (\citealp{Patel}). To assess the
effect of ambiguous classifications in our study we will also, next to
the full sample, consider the subset of Herbig Ae/Be stars in Table 1
of \citet{The}. This catalogue contains all historically known, and
best studied, Herbig Ae/Be stars. {98/254} of our
{initial} sources {with parallaxes} are present in this table (their best
candidates).


\subsection{Atmospheric parameters, photometry and extinction values}\label{sec:Atmospheric}

\begin{figure*}[ht!]
\centering\includegraphics{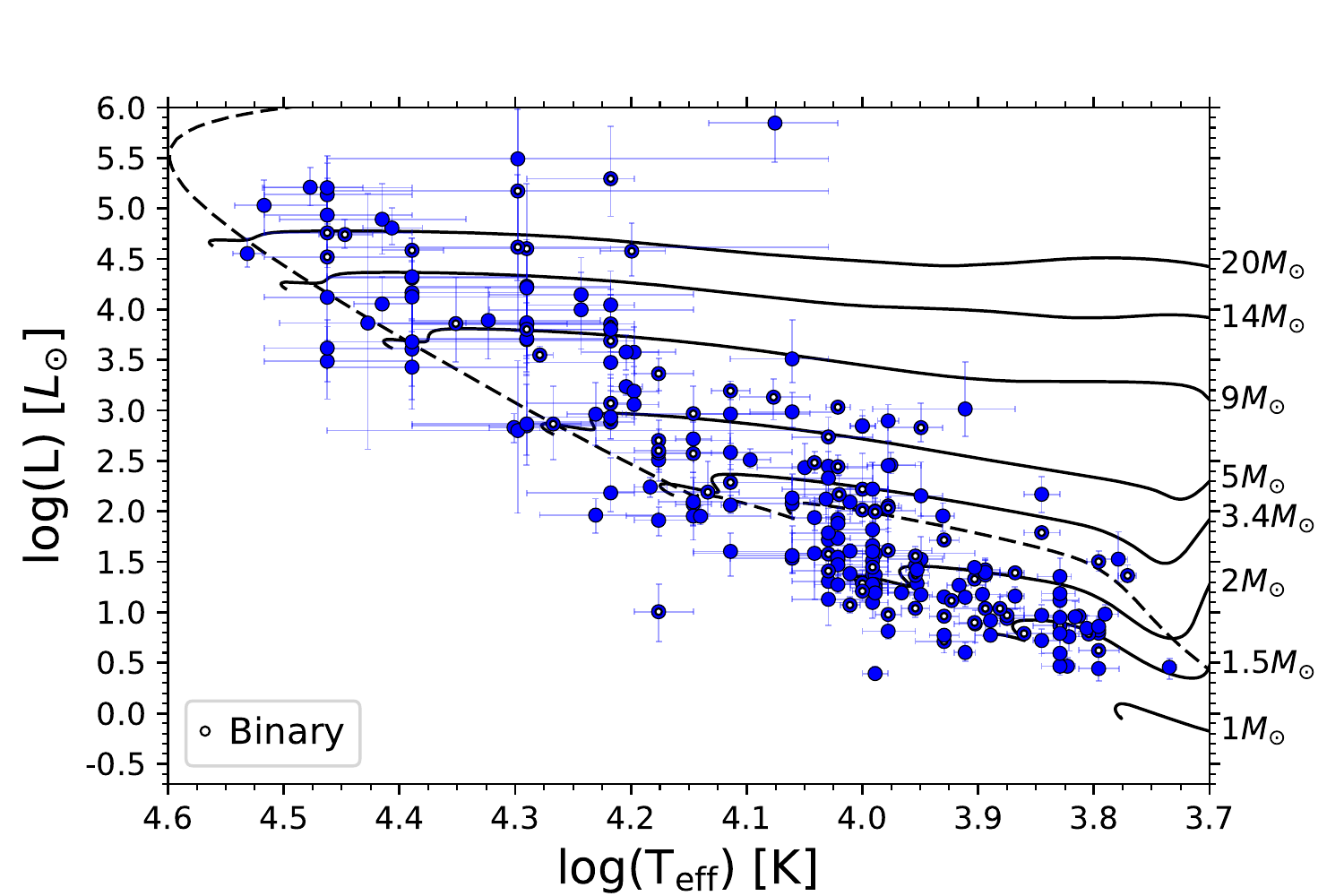}
\caption{{223} Herbig Ae/Be stars in the HR diagram {satisfying Eq.~\ref{astrom} constraint}. 
{In most cases vertical error bars are dominated by parallax uncertainties}. Sources with a white dot
  have been classified as binaries. The mass of each Pre-Main Sequence
  track \citep{Bressan} is indicated on the righthand side. {An isochrone (\citealp{Bressan};
\citealp{Marigo}) of $2.5$ Myr is also shown for reference as a dashed line.}\label{hrd}}
\end{figure*} 

\begin{figure*}[ht!]
\centering\includegraphics[scale=0.61]{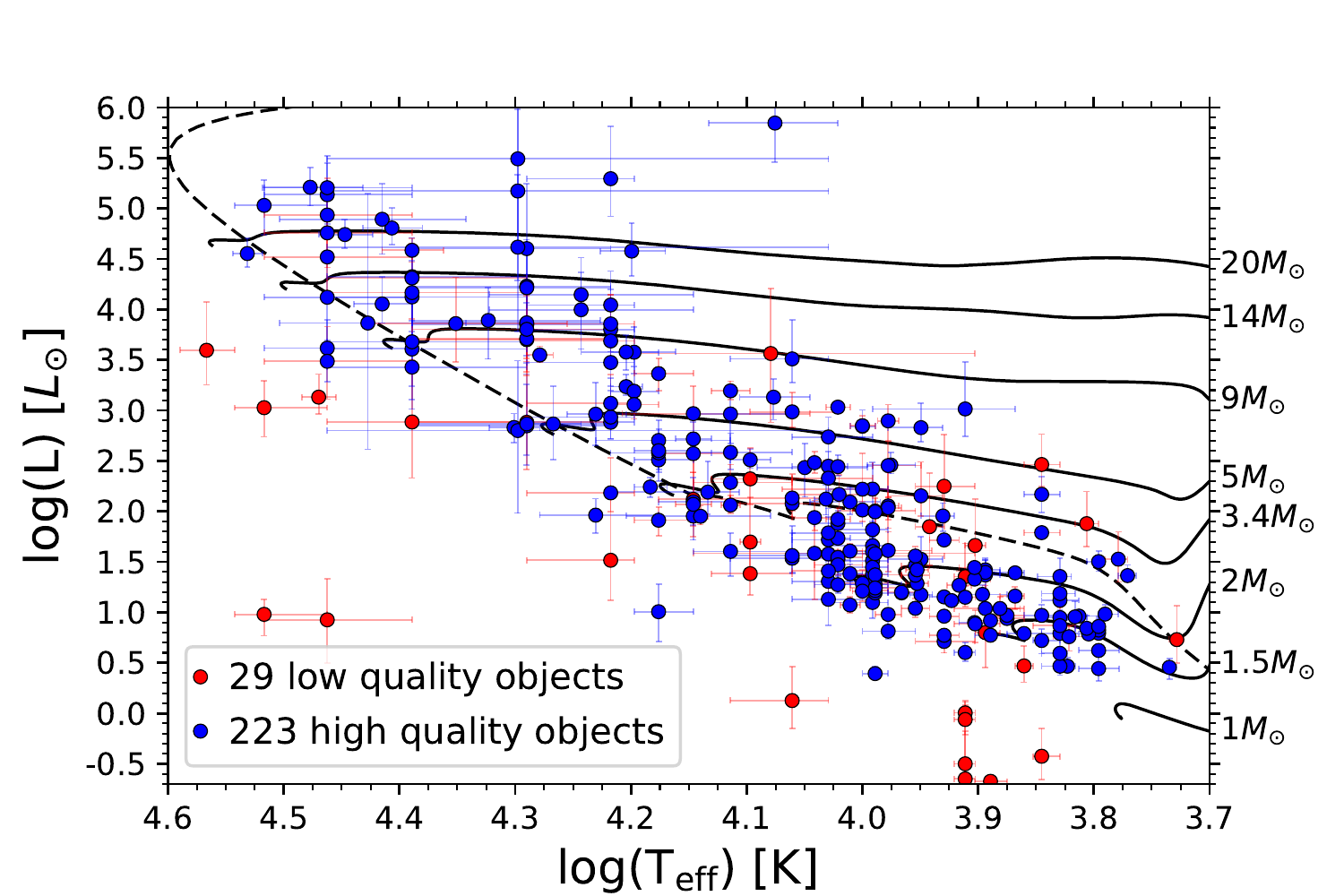}\includegraphics[scale=0.61]{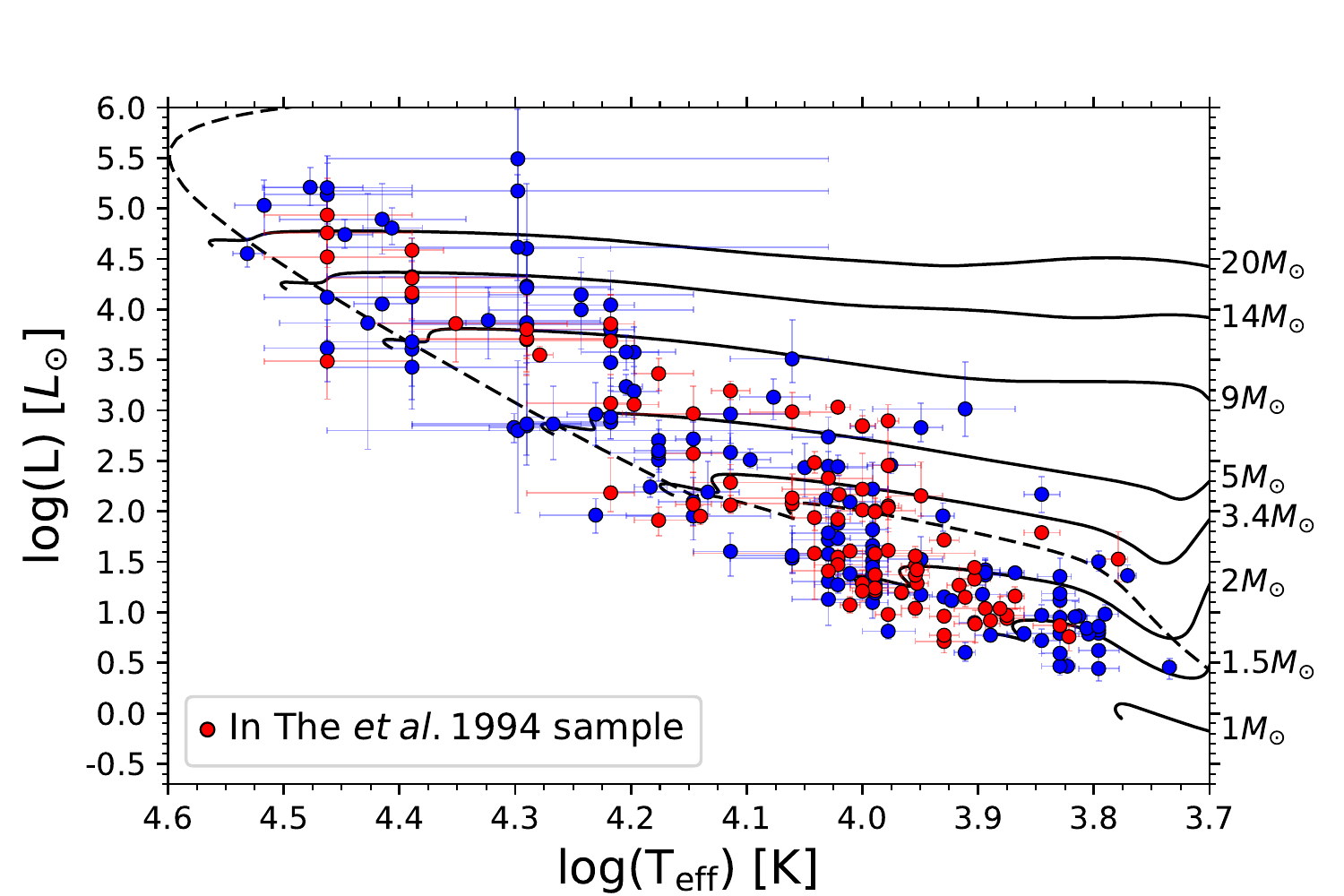}
\caption{Left: {223 high quality and 29 low quality Herbig Ae/Be stars in the HR diagram after the cut in astrometric quality described in Eq.~\ref{astrom}. Right: 218 Herbig Ae/Be stars in the final high quality sample after removing the 5 problematic objects described in Sect. \ref{sec:HRdiagram}. In red those objects present in Table 1 of \citet{The}. The mass of each Pre-Main Sequence
  track \citep{Bressan} is indicated on the righthand side. {An isochrone (\citealp{Bressan};
\citealp{Marigo}) of $2.5$ Myr is also shown for reference as a dashed line.}\label{hrd2}}}
\end{figure*} 

We obtained atmospheric parameters and photometric and extinction
values for all the sources from the literature. These were mainly
{\citealp{Alecian}; \citealp{Carmona}; \citealp{Chen};
  \citealp{Fairlamb}; \citealp{Hernandez}; \citealp{Hernandez2};
  \citealp{Manoj}; \citealp{Montesinos}; \citealp{Mendigutia};
  \citealp{Sartori}; \citealp{Vieira} and the APASS Data Release
  9}. Whenever the effective temperature (T\textsubscript{eff}) was
not available it was derived from the spectral type with the effective
temperature calibration tables of \cite{Gray}. A 1 subspectral type
uncertainty was assigned in all cases. When not listed in the
literature, A\textsubscript{V} values were derived from the observed
photometry and using the intrinsic colours of \cite{Pecaut}. 
An $R_V=3.1$ was used in all cases in which
A\textsubscript{V} was derived although other studies like
\citet{Hernandez} or \citet{Manoj} have suggested that a larger value
of for example $R_V=5$ could be more appropriate for HAeBes where
local extinction dominates the total extinction. This is a topic for
future investigations using Diffuse Interstellar Bands (as done by for
example \citealp{Oudmaijer6}). The relevant data of each source is
presented in {Table~\ref{table2} and Table~\ref{table2_2} for the high
quality and low quality samples respectively.}

{HAeBes usually show photometric variability.}
Thus, for objects with multi-epoch photometry available, 
we selected the brightest set to determine the extinction
towards the objects and thus their intrinsic brightnesses.  As we will
also show later, the variability is often caused by irregular
extinction, using those data with minimum extinction introduces the
smallest errors in the determination of the stellar parameters.
{For this reason, we only used simultaneous photometry when 
deriving A\textsubscript{V} values.}
All the photometric values were corrected for extinction 
using the reddening law of \citet{Cardelli}.

{Two sources, \object{V833 Ori} and \object{GSC 1829-0331}, do not have enough
simultaneous photometry available to derive extinctions for them and hence
they were excluded for the sample. The total number of Herbig Ae/Be stars that
can be placed in the HR diagram and for which we can derive stellar luminosities, masses, ages, IR excesses and variabilities in Sect. \ref{sec:Der_cuantities} is therefore reduced to 252 objects.}

\subsection{Infrared photometry}

All the sources were crossmatched with the Two-Micron All Sky Survey
(2MASS, see \citealp{2MASS}) and with the \textit{Wide-Field Infrared
  Survey Explorer `AllWISE'} all-sky catalogue (hereafter WISE, see
\citealp{ALLWISE}). Both these surveys contain hundreds of millions of
stars, guaranteeing a large overlap with Gaia.
We used a 3 arcsecond aperture for the crossmatch.
The few sources that did not lie within that 3 arcsecond threshold were
studied individually and, if present, their IR photometry was
included. This provides values and uncertainties for the J, H
and K\textsubscript{s} bands ($1.24$, $1.66$ and $2.16\mu$m
respectively) and for the W1, W2, W3 and W4 bands ($3.4$, $4.6$, $12$,
and $22\mu$m respectively) for most of the HAeBes. Note that for some
sources some of the bands may be missing or just be upper limits.  
We double-checked all infrared matches with the dereddened optical photometry
and found no inconsistencies.

\subsection{H$\alpha$ equivalent width and emission line profile}\label{sec:EWandLine}

We collected all the H$\alpha$ equivalent widths (EW) we could find in
the literature for the Herbig Ae/Be stars. Not only the intensity of
the line but also the shape contains very useful
information. Therefore, when possible, information about the shape of
the H$\alpha$ line was included. We have classified the H$\alpha$ line
profile as single-peaked (s), double-peaked (d) and showing a P-Cygni
profile (P), both regular or inverse.  EW and line shape information
are presented in {Table~\ref{table3} and Table~\ref{table3_2}
  for the high quality and low quality samples respectively.}  Many
Herbig Ae/Be stars are quite variable in their H$\alpha$ emission and
its EW may significantly change at short timescales (\textit{e.g.}
\citealp{Costigan}). This is also the case for the line shape,
although spot checks on objects that have more than one H$\alpha$
observation listed in the literature appear to indicate that there are
not many changes in line profile classification (see also for example
\citealp{Aarnio}), although changes between single-peaked and double-peaked 
profiles in a given star are also observed
(\citealp{Mendigutia3}). We do note the additional complication that
emission line shapes are often difficult to unambiguously classify.

Regarding the H$\alpha$ EWs compiled, we note that our main references
(\citealp{Fairlamb} and \citealp{Mendigutia3}) provide the
non-photospheric contribution of the EW, while most other authors
state the observed EW, which includes the photospheric contribution.
This photospheric absorption peaks for A0-A1 type objects, with EW
values of $\sim+10\AA$ (See \textit{e.g.} Fig. 7 of \citealp{Joner}) but is
only $\sim+2\AA$ for B0 objects. {We used the \citet{Joner}
  results to correct those EWs that were not corrected for
  absorption.}

We have H$\alpha$ EWs for {218/252} of the HAeBes and line
profiles for {197} of these: {31\%} are single-peaked,
{52\%} double-peaked and {17\%} P-Cygni (of which the
vast majority are of regular P-Cygni type). This is in agreement with
\citet{Finkenzeller} who found that out of 57 HAeBes, 25\% were
single-peaked, 50\% showed double-peaked H$\alpha$ profiles and 20\%
presented a P-Cygni profile (both regular and inverse). The main
references for the EW values are \citet{Baines}; \citet{Fairlamb2}; \citet{Hernandez}; 
\citet{Mendigutia3} and \citet{Wheelwright}. Main references for the line profiles are \citet{van den Ancker3}; \citet{Baines}; \citet{Mendigutia3}; \citet{Vieira} and
\citet{Wheelwright}. The rest of the
references can be found in {Table~\ref{table3} and
  Table~\ref{table3_2}.}

\subsection{Binarity}

More than half of the Herbig Ae/Be stars are known to be in binary
systems (\citealp{Duchene}). The true number is likely much larger, as
there have been a small number of targeted surveys for binarity of
HAeBe stars, the largest are \citet{Wheelwright} and \citet{Baines}
who performed spectro-astromety of 45 HAeBes and 31 HAeBes
respectively, probing companions in the $\approx0.1$ - $2$ arcsec
range,
and \citet{Leinert} who performed speckle interferometry of 31
objects, sampling separations of order $0.1$ arcsec.  {81/252
  HAeBes ($\sim32\%$)} of our set are catalogued as binary systems, a
fraction in agreement with the \citet{Duchene} findings {if we take
  into account the large number of faint Herbig Ae/Be stars which have never been
  studied for binarity}.  The binary status of each HAeBe is presented
in {Table~\ref{table2} and Table~\ref{table2_2} for the high quality and low quality samples respectively}; main references were \citet{Baines};
\citet{Leinert} and \citet{Wheelwright}, we refer to
{Table~\ref{table2} and Table~\ref{table2_2}} for a complete list.

\citet{Baines} found a typical wide (few hundred au) separation in the
binary systems. \citet{Wheelwright} detected no binaries closer than
30~au and established a range of $\approx40$ - $4000$~au in their data.

\section{Derived quantities}\label{sec:Der_cuantities}

\subsection{Luminosity and Hertzsprung-Russell diagram}\label{sec:HRdiagram}

Using the parallaxes, atmospheric parameters and extinction values we
derived the luminosity for the {252 HAeBes} with parallaxes
employing a similar method to \citet{Fairlamb}, which is similar to
\citet{Montesinos} and \citet{van den Ancker2}. In short, it first
consists of using values of T\textsubscript{eff} and surface gravity
(log(g)) to select an atmosphere model from \citet{Castelli} (referred
to as CK-models hereafter) for each star to be used for its intrinsic
spectral energy distribution (SED).  Solar metallicity CK-models were
used in all cases but for \object{BF Ori}, \object{RR Tau}, \object{SV
  Cep}, \object{XY Per} and \object{WW Vul} for which the
metallicities are known not to be solar from the spectroscopic work of
\citet{Montesinos}.  When possible, the log(g) values were estimated
from the luminosity class; otherwise they were taken as 4.00 (typical
values range from 3.5 to 4.5). Uncertainties in log(g) and metallicity
can be neglected in our study as their effect on the model SED and
derived quantities is negligible.

We then scaled the model to the dereddened photometric Johnson V
band.  The energy distribution is then integrated over frequency to
get the total flux. The final luminosities, presented in
{Table~\ref{table2} and Table~\ref{table2_2} for the high quality and low quality samples respectively}, are then obtained by means of the total flux and
the parallax.  All sources of uncertainty were taken into account at
this step {including} using different CK models for the
different temperatures within the $T\textsubscript{eff}$ uncertainty range.


The {223 
 Herbig Ae/Be stars {satisfying Eq.~\ref{astrom} constraint} are plotted in
the resulting HR diagram in Fig.~\ref{hrd}. This number is an increase
of more than a factor of ten compared to the previous,
Hipparcos-based, study by \citet{van den Ancker3}.  PMS evolutionary
tracks from \citet{Bressan} are also plotted in Fig.~\ref{hrd} plus a $2.5$Myr
isochrone (\citealp{Bressan}, \citealp{Marigo}), all of them with solar metallicities
($Z=0.01$ and $Y=0.267$).}

{Before we analyze this sample, we also plot the HR diagram for
  all 252 objects with parallaxes (the high and low quality samples together,
  hence including those which failed the
  Lindegren quality selection criteria) in Fig.~\ref{hrd2} on the left.
  Many of these badly astrometrically behaved sources are located in unphysical positions,
  significantly below the Main Sequence, validating our approach of
  removing those from our analyses.}

{Returning to the HR diagram in Fig.~\ref{hrd}, there are still several
  outliers that do not seem to be PMS objects. \object{GSC 5360-1033} and \object{UY Ori}
  appear way below the Main Sequence, just like the lower quality
  objects that were removed earlier. However, the Gaia DR2 data of
  these two objects appear of good quality. 
  Regarding GSC 5360-1033 the
  situation is not clear, an ambiguous spectral type or photometry for this object, 
  or an incorrect estimation or the extinction may be the
  reason for the unexpected location of the object. 
  For UY Ori, Fairlamb et al. 2015 assigned a spectral
  type of B9 to this object, but the photometry listed in {\sc simbad}
  indicates a large variability.
  Pending more certainty, we decided to
  remove these two objects from the high quality sample 
  and place them in the low quality sample.}

  {\object{MWC 314}, \object{MWC 623} and \object{MWC 930} on the other hand appear quite luminous
  and very much to the right of the Main Sequence, something very unusual
  for high mass PMS objects. An individual inspection reveals that these
  objects are more likely to be evolved giants and they appear in the literature
  as such (\textit{e.g.} for \object{MWC 314}: \citealp{Carmona}, for \object{MWC 623}: \citealp{Lee}, for \object{MWC 930}: \citealp{Miroshnichenko_LBV}).}  
   {Deciding on the
  nature of the various Herbig Ae/Be candidates in our master sample
  is not our intention and it is beyond the scope of this paper that is
  essentially a statistical study. However, these objects occupy a special place in the HR diagram that is consistent with both a pre- as well as a post-Main Sequence nature while there is much information regarding these objects supporting their post-Main Sequence nature. We therefore decided to err on the cautious side and
  remove these as well from further analysis.}

  {The final HR diagram without these $2+3$
  problematic objects is presented in Fig.~\ref{hrd2} on the right. In addition, in this graph, we highlight the sample of The et al. 1994 bonafide HAeBes in red. This final
  high quality sample of 218 objects will be the one we will use in
  the following plots and studies.  The information concerning the 34
  discarded objects in the low quality sample can be found in the tables at the end of the
  paper.}

In this last high quality HR diagram we see that there are many more
low mass HAeBes than high mass HAeBes ({69\%} of the sources
are below $4M_{\odot}$). This is most likely because of the Initial
Mass Function (IMF). This trend of more objects for lower masses
discontinues below $\sim2 M_{\odot}$. This is roughly the mass
corresponding to the boundary between Main Sequence A- and F-type
stars, and thus the traditional lower mass boundary at which the
Herbig Ae/Be stars were originally selected.

For lower masses the sources are more spread in temperature, occupying
larger parts of the PMS tracks, while, instead, the high mass objects
tend to be predominately located close to the Zero Age Main Sequence
(ZAMS). This is likely because the higher the mass the faster the PMS
evolution is. This fast evolution could explain why high
mass objects at low temperatures (and thus low surface gravities) are
hardly present in Fig.~\ref{hrd} or the sample.

We will encounter more examples below where high mass and low mass
objects have different properties.

\subsection{Mass and age}

Using the isochrones, the mass and age of the Herbig Ae/Be stars were
estimated. We used a hundred PARSEC isochrones with solar metallicity (\citealp{Bressan};
\citealp{Marigo}) from $0.01$ to $20$Myr, from which we only use the
PMS tracks. To each Herbig Ae/Be star we assigned the closest two
isochrone points in the HR diagram; {the solar metallicity isochrones 
did not match seven sources from the high quality sample in the HR diagram and isochrones with lower metallicities were used in those cases}. 
As each point is associated with a
mass ($M$) and an age, we computed for each HAeBe an average of those
values weighted by the distance to the points. The result is an
estimate of age and mass for {236/252} HAeBes. These values are
presented in {Table~\ref{table3} and Table~\ref{table3_2} for the high quality and low quality samples respectively}. Uncertainties were derived from the
error bars in the HR diagram (Fig.~\ref{hrd} and Fig.~\ref{hrd2}) {keeping a minimum error of $5\%$.} 
We compared many of
our masses and ages with those of \citet{Alecian} and
\citet{Reiter}. We found that our determinations of these parameters
are consistent with the results of the previous authors.

\subsection{Infrared excesses}

In the process of deriving the luminosity it is also possible to
derive the infrared excess. We have logarithmically interpolated the
different dereddened observed fluxes from the J band ($1.24\mu$m) to
the W4 band ($22\mu$m) and defined the infrared excess ($E$) as:

\begin{equation}\label{eq1}
E=\frac{(F\textsubscript{e}-F_{*})_{[\lambda_{1},\lambda_{2}]}}{F_{*}}
\end{equation}

$F\textsubscript{e}$ is the total flux {underneath} the \textit{observed
  dereddened photometry} {(the infrared photometry has also
  been dereddened)} and $F_{*}$ is the {total} photospheric
flux below the CK model. {$\lambda_{1}$ and $\lambda_{2}$
  define the range of wavelengths of interest and the total fluxes in
  the numerator just refer to that range.} This measure expresses the
excess in terms of the total luminosity of the object. For example,
all things being equal, if we have two stars with the same amount of
dust surrounding them, with one of them brighter, the infrared
re-radiated emission will be larger, but the IR excess as defined
here, would be the same as it is a relative measure. The same or a
very similar indicator was used by \citet[their Eq. 8]{Cote},
\citet[Eq. 3]{Waters}, and more recently by \citet{Banzatti} in their
Sect. 2.3.

Uncertainties in the infrared excesses were derived using the uncertainties in the 
observed fluxes and the uncertainties in the temperature 
(which affect the CK models) of each object.

We have split the total infrared excess in two, a Near Infrared
Excess ($1.24-3.4\mu$m, roughly the 2MASS region) and a Mid Infrared
Excess ($3.4-22\mu$m, the WISE region). {The values for these excesses
are presented in Table~\ref{table3} and Table~\ref{table3_2} for the 
high quality and low quality samples respectively}. The total infrared excess ($1.24-22\mu$m)
is the sum of the two.

{In addition, we also computed the infrared excess at each individual band 
(J, H, K\textsubscript{s}, W1, W2, W3 and W4) as the flux ratio between the dereddened observed monochromatic flux
and the expected flux according to the CK model.
 The values for these excesses
are presented in Table~\ref{table4} and Table~\ref{table4_2} for the 
high quality and low quality samples respectively}

\subsection{Variability information}\label{sec:Variability}

Gaia DR2 does not provide a general variability indicator {for all sources}.
Here, we use Gaia's repeated observations to extract photometric variability
information. Gaia DR2 used a total of 22 months of observations and each source
 was observed repeatedly in a non periodic fashion. Data Release 2 provides the average photometry, the
uncertainty on this value and the number of observations. All things
being equal, the photometric ``error'' will be larger for a
photometrically variable object than for a stable object. Here we aim
to quantify the variability of the objects.  We start with the
``Variability Amplitude'' ($A_{i}$) for a certain source \textit{i} as
presented in \citet{Deason}:

\begin{equation} 
A_{i}=\sqrt{N\textsubscript{obs,\textit{i}}}\;e(F_{i})/F_{i}
\label{eq2}\end{equation}

where $N\textsubscript{obs}$ is the number of CCD crossings, $F$ and $e(F)$ are the
flux and flux error respectively. This quantity is powerful in
identifying objects that show larger flux variations than expected for
a stable star. However, in order to statistically assess the level of
variability, we introduce a variability indicator $V_{i}$, which
quantifies how much more variable an object is compared to stable
objects of the same brightness. In short it compares the Variability
Amplitude from Eq.~\ref{eq2} of a given object ($i$) to that of all Gaia
objects in a brightness interval of $\pm$0.1 magnitude around the G
band value of the object (\textit{i.e.} to
${A_{a}}_{,G_{a}\in{(a_{1},a_{2})}}$, with \textit{a} indexing the
Gaia catalogue and being $a_{1}=G_{i}-0.1$mag and
$a_{2}=G_{i}+0.1$mag). The equation is as follows:

\begin{equation} 
V_{i}=\frac{A_{i}-\overline{A_{a}}_{,G_{a}\in{(a_{1},a_{2})}}}{\sigma[A_{a}]_{G_{a}\in{(a_{1},a_{2})}}} 
\label{eq3}\end{equation}

$G$ is the Gaia white G band magnitude and $\sigma$ is the
standard deviation.
In essence, we subtract the error to flux ratio of each HAeBe,
weighted by the number of observations, to the mean of the same
expression ($A_{a}$, Eq.~\ref{eq2}) for the sources in the Gaia catalogue
within $\pm{\rm 0.1mag}$ of the Herbig star in the G band. Then we
divide by the standard deviation of $A_{a}s$ of that Gaia subset.
This results in a variability indicator which measures the variability
(in standard deviations, $\sigma$) for each Herbig Ae/Be star compared to the
mean of field objects of the same brightness.

\begin{figure}
\includegraphics[scale=0.60]{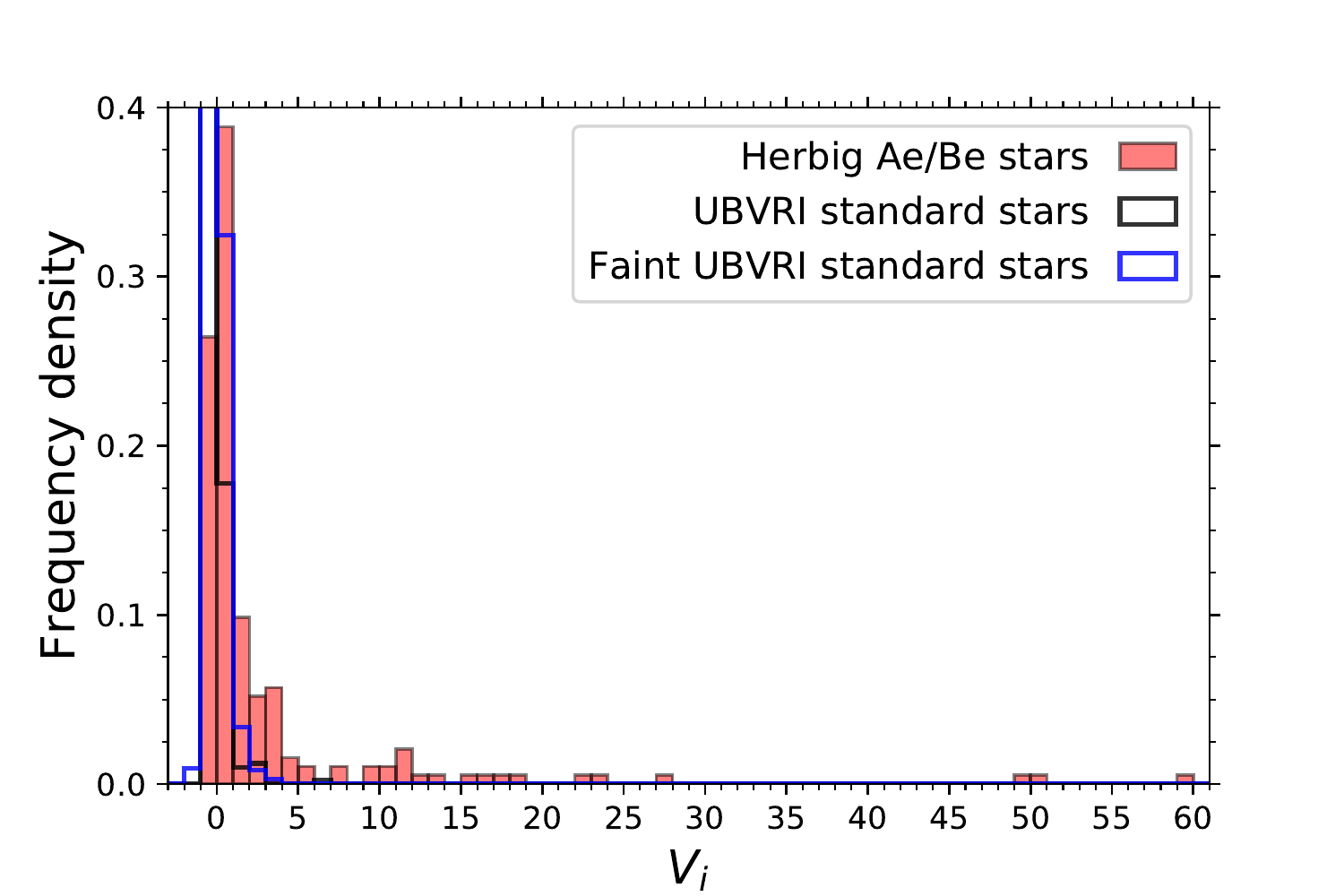}
\caption{Distribution of the variability indicator for Herbig Ae/Be
  stars and two catalogues of photometric standards; one of bright
  sources (\citealp{Landolt}) and one of faint sources
  (\citealp{Clem}). As a class, the Herbig Ae/Be stars are more
  variable than the photometric standards.  \label{Histogram}}
\end{figure}

\begin{figure*}
\centering\includegraphics[scale=0.58]{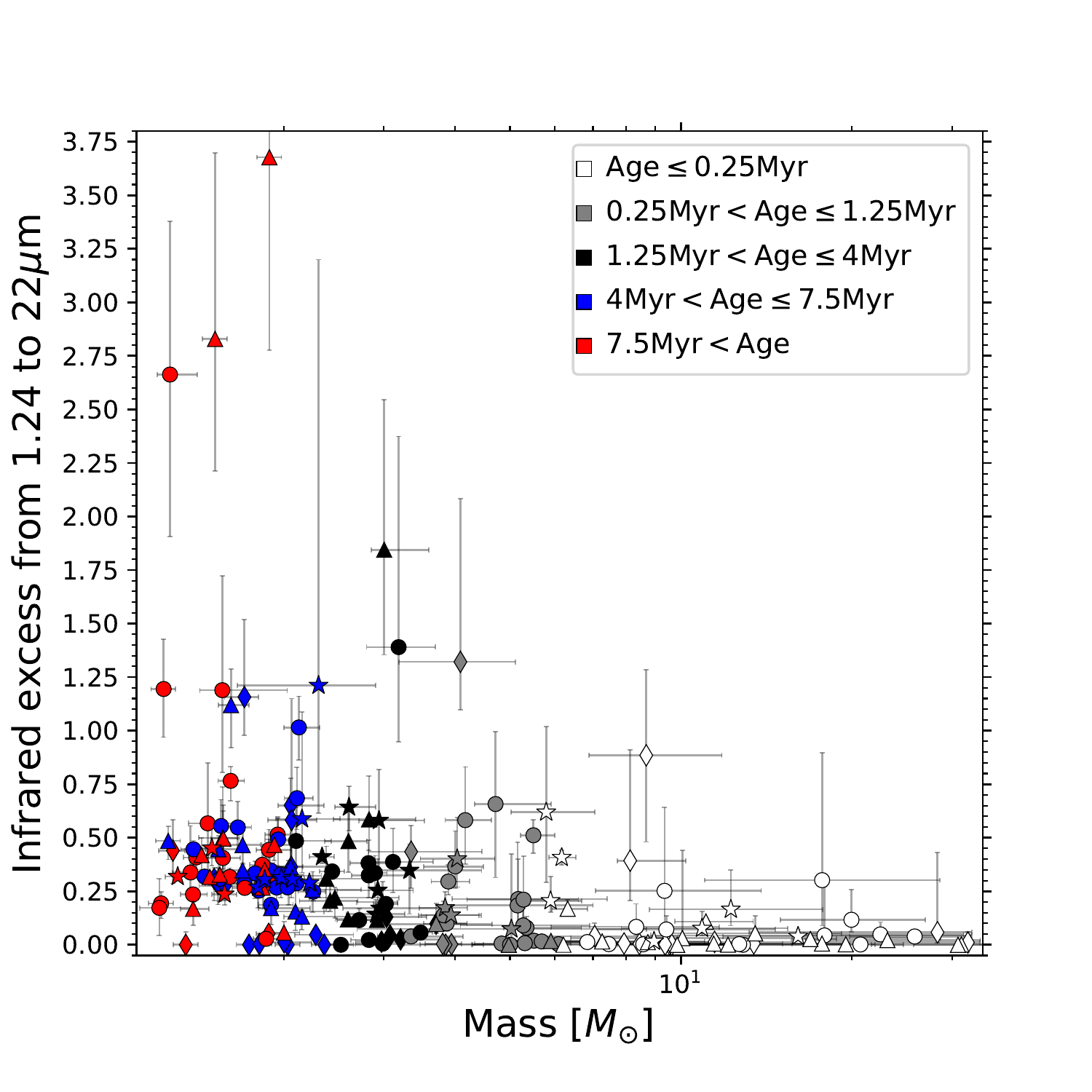}
\includegraphics[scale=0.58]{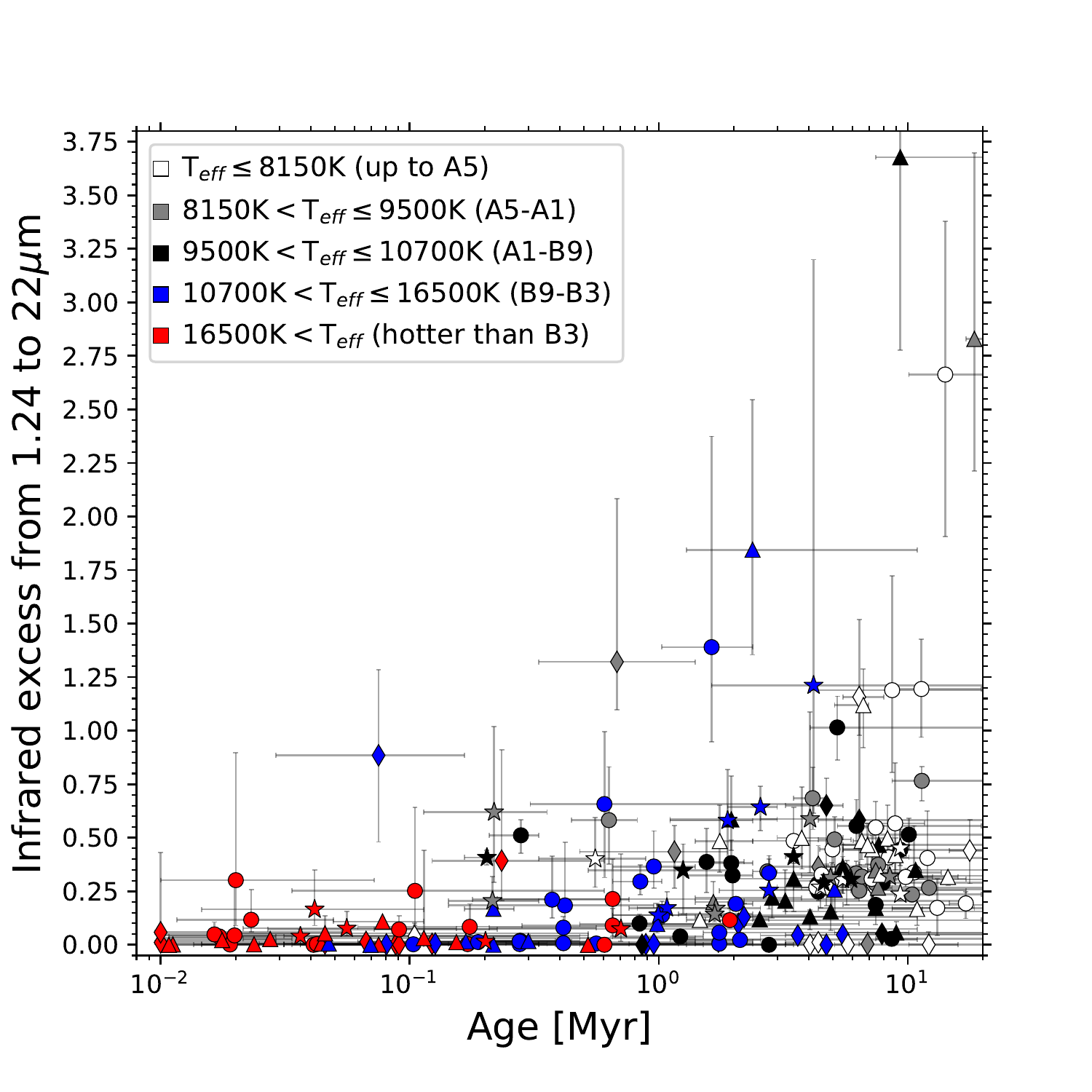}

\caption{Left: IR excess in the range $1.24-22\mu$m vs. estimated mass of the
  objects. The most massive objects (more massive
  than $\sim7 M_{\odot}$)  barely show an infrared excess. Right:
  IR excess in the range $1.24-22\mu$m vs. estimated age. Ages and
  effective temperatures are respectively colour coded in the legend.  The symbols stand
  for the H$\alpha$ line profiles: circles (double-peaked), triangles
  (single-peaked), stars (P-Cygni profile) and diamonds (no
  information).  Note that although it is not necessarily a one-to-one
  correlation, lower ages correspond to higher masses.}\label{irvsm}
\end{figure*}

For completeness, we note that it is necessary to impose more
constraints to exclude the cases in which a larger error is not due to
intrinsic variability. Following \citet{Deason}, {Appendix C of \citet{Lindegren_new} and
what was done in \citet{HRdiagram}} we require
$N\textsubscript{obs}\geqslant70$
{and more than 8 visibility periods (\textit{i.e.}, groups of observations separated by at least four days), 
plus the Eq.~\ref{astrom} constraint that limits the astrometric quality (and hence 
the variability indicator just can be
derived for sources in the high quality sample).
In order to also limit the photometric quality
we included the following criterion presented in \citet{HRdiagram}}:

\begin{equation} 
1.0+0.015(G\textsubscript{BP}-G\textsubscript{RP})^2<E\textsubscript{F}<1.3+0.06(G\textsubscript{BP}-G\textsubscript{RP})^2
\label{phot}\end{equation}

{where E\textsubscript{F} is the flux excess factor and $G\textsubscript{BP}$ and $G\textsubscript{RP}$ 
the Gaia blue and red passbands respectively. 
Note that these constraints may inevitably exclude
many of the very variable HAeBes as they also trace larger errors and hence variability.
These constraints will also be biased to discarding binaries
and faint sources in crowded areas (\citealp{Lindegren_new}; \citealp{HRdiagram}).}

The variability indicator values for the {193} sources satisfying the previous conditions are 
presented in Table~\ref{table3}. 

In Fig.~\ref{Histogram}, we show the $V_{i}$ distribution of Herbig
Ae/Be stars and compare it to the $V_{i}$ distribution of bright
photometric standards from \citet{Landolt}
and faint photometric standards taken from \citet{Clem}. If Eq.~\ref{eq3} would have not been
used these two latter samples would have had a different mean in the
distribution of $A_{i}$. The Herbig Ae/Be stars appear to show, on
average, a larger variability indicator value than the standard stars,
being the break at $\sim V_{i}=2$. We performed a two-sample
Kolmogorov-Smirnov (KS) statistical test to study whether Herbig Ae/Be
stars can be drawn from those two samples of standard stars. The
result shows that we can reject that hypothesis to within a 0.001
significance and hence this variability indicator differentiates them
as a group.

In order to assess the relation between our variability indicator (G band variability) and
variability in the V band we compared the magnitude variations in the
V band as presented in the International Variable Star Index VSX
(\citealp{Watson}) with our variability indicator values. We found that we are
tracing variabilities as small as $\sim0.5$mag with the $V_{i}=2$
cut-off. In \citet{Eiroa} 7/23 (30\%) PMS objects homogeneously
observed for variability have variabilities above 0.5mag. In our case
{48/193} sources have values above $V_{i}=2$ {(25\%)} and hence can be
considered as strongly variable. Of those {48, 17} are catalogued as
UXOR type (\citealp{Mendigutia_thesis}; \citealp{Oudmaijer5}; \citealp{Poxon}). 
{There are 5 other UXORs in our sample with $V_{i}$ values, 4 of them have reported optical variabilities
smaller than $0.5$mag in the V band. The other one is \object{BO Cep}. This object has been
reported to have a periodic variability with a single prominent peak with
a period of $\sim10$ days (\citealp{Gurtler}). The regular non periodic variability of the object
is smaller than $0.5$mag which explains why this UXOR has not been detected by our
variability indicator. Supporting this, it appears as  UXOR in \citet{Poxon} but not in 
\citet{Oudmaijer5} or \citet{Mendigutia_thesis}.}

To put the variability indicator into perspective, we find that {6 out of 411} photometric
standards from \citet{Landolt} have variability indicator values larger than 2. 
We would therefore expect only {3 of our 193} Herbig Ae/Be stars for
which we could determine $V_{i}$ to be strongly variable, at amplitudes of 0.5
magnitudes in the V band or higher. However, we find {45} more, indicating that a large fraction of Herbig Ae/Be stars exhibit  strong variations.
 
{In addition, it is interesting to compare our variability
  indicator with the variability catalogues published alongside the
  Gaia DR2 general catalogue (\citealp{Holl}).  Just 1 every 3000
  objects passed the Gaia DR2 stringent selection criteria for variability.
  10/252 objects in our
  list fall in this category and appear as variable in those catalogues.  Of the 5 of those that
  have derived $V_{i}$ values they are larger than $V_{i}=5$.}
\section{Data analysis}\label{sec:Data_analysis}

\subsection{Infrared excesses}\label{sec:IR_excess}

In Fig.~\ref{irvsm} the total infrared excess ($1.24-22\mu$m) vs. the
estimated mass and age of the sources is plotted.  There appears a
difference in IR properties between high and low mass stars. Whereas
low mass stars show a range of infrared excess, the higher mass stars
{in general} only present very low levels of excess. A similar behaviour is seen
when the excess is plotted as function of age; the excess for the
youngest objects is smallest. This is probably readily explained by
the fact that the more massive PMS objects in the HR diagram have the
lowest ages by virtue of their rapidly evolving isochrones, so trends
in mass will automatically also be present in those with age. To study
trends as a function of age, it would be necessary to consider subsamples with a narrow
range in mass.  We therefore consider that the main result of this
exercise is that high mass objects have a very low infrared excess,
and that there appears to be a break at $\sim7 M_{\odot}$ from where {almost} no
sources with significant excess appear.

\begin{figure}[t]
  \begin{center}
  \includegraphics[scale=0.62]{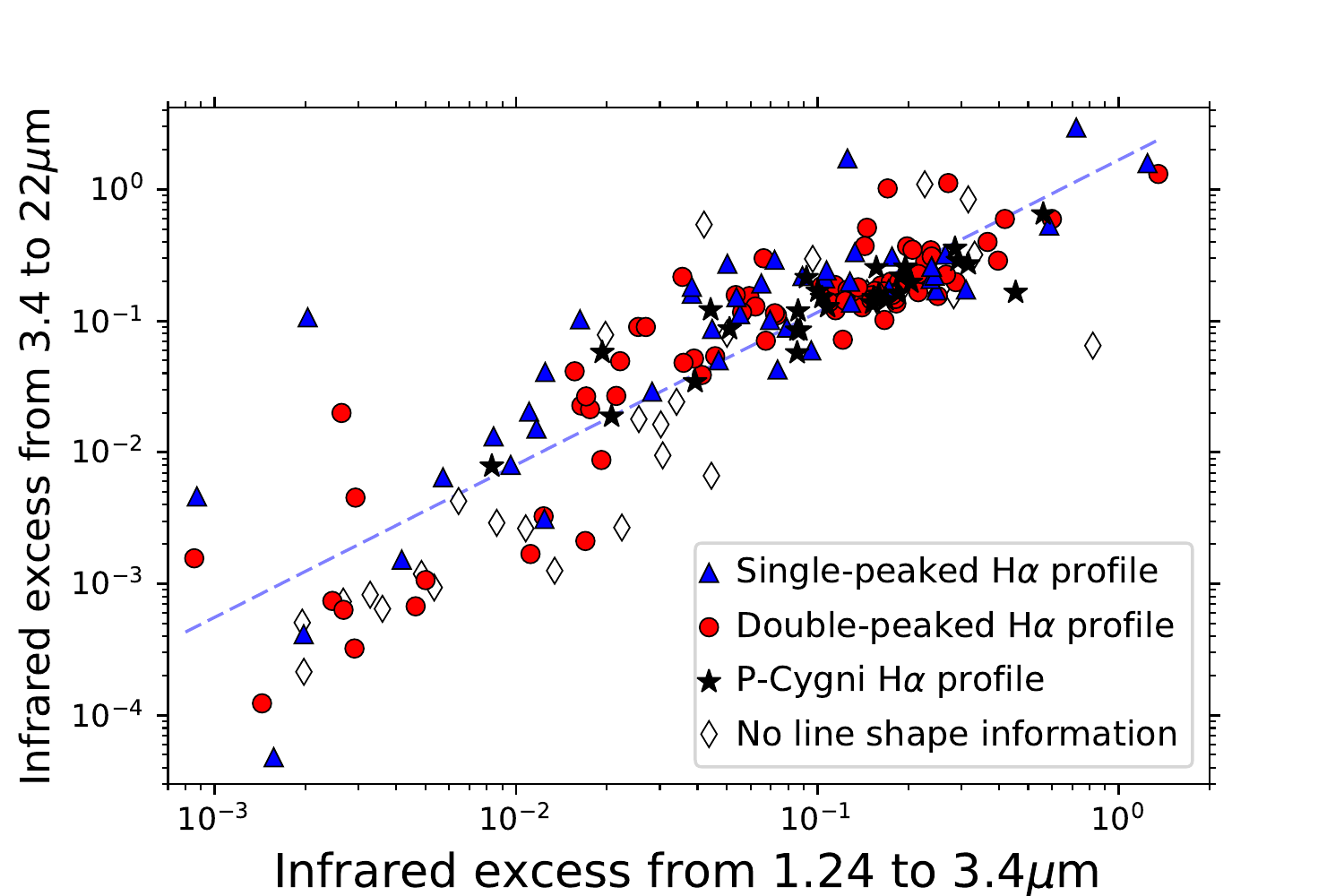}
\caption{IR excess in the range $3.4-22\mu$m (Mid IR excess) vs. IR excess in the range $1.24-3.4\mu$m (Near IR excess). The symbols stand
  for the H$\alpha$ line profiles: circles (double-peaked), triangles
  (single-peaked), stars (P-Cygni profile) and diamonds (no
  information). A linear fit in the log space is shown in blue {($\log($Mid IR\textsubscript{excess}$)=1.16\log($Near IR\textsubscript{excess}$)+0.23$, $r=0.88$)}. }\label{irvsir}
  \end{center}
\end{figure}

Fig.~\ref{irvsir} splits the total infrared excess into two, a
near-infrared and a mid-infrared part. It demonstrates that the excess
at both wavelength ranges are highly correlated with each other (the
linear fit in logarithmic space that can be seen in the plot has a
correlation coefficient of {$r=0.88$}). Therefore, it is not unexpected 
that the $\sim7 M_{\odot}$ break is also present at near- and mid-infrared.

\subsection{H$\alpha$ equivalent width}

\begin{figure*}[ht!]
\centering\includegraphics[scale=0.58]{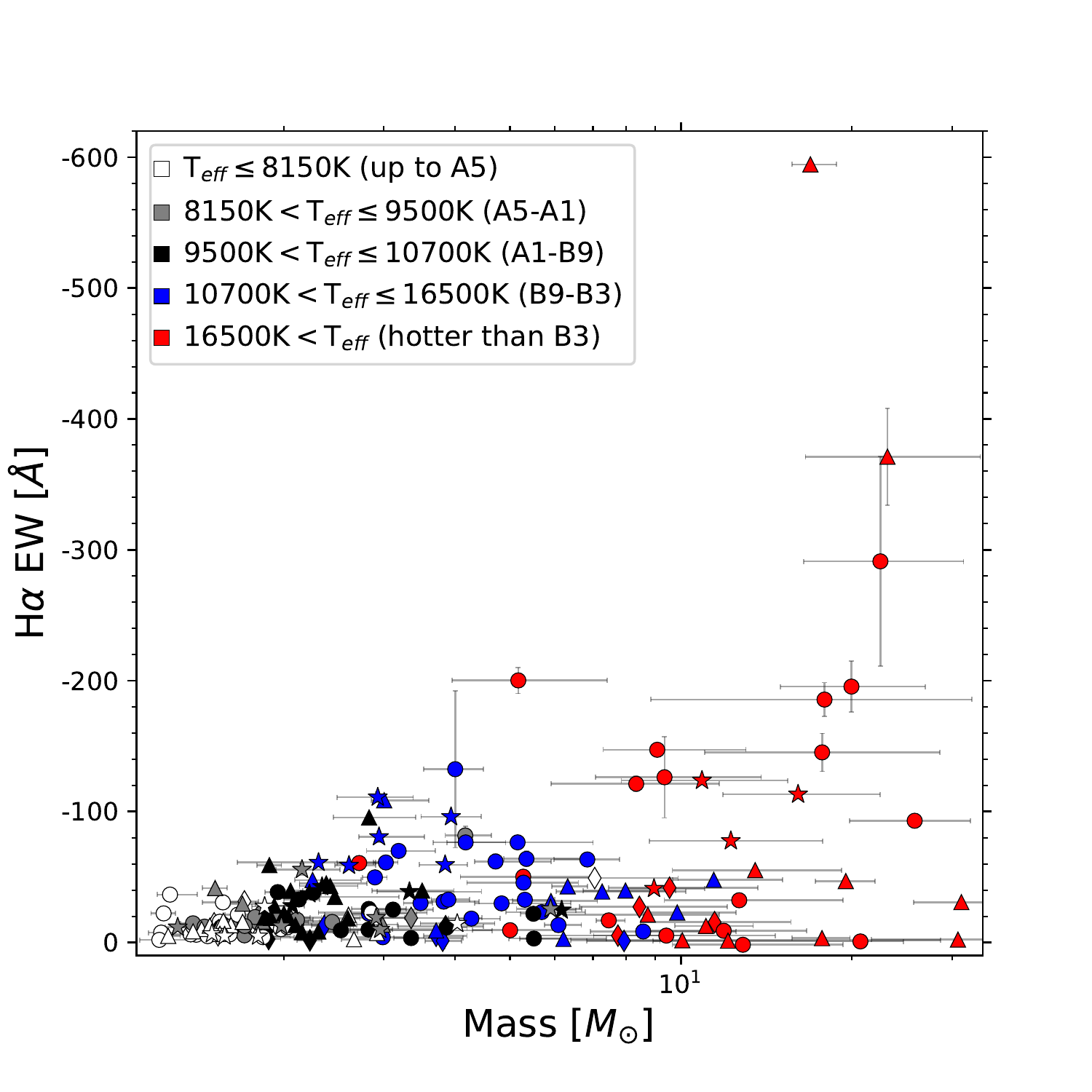}
\includegraphics[scale=0.58]{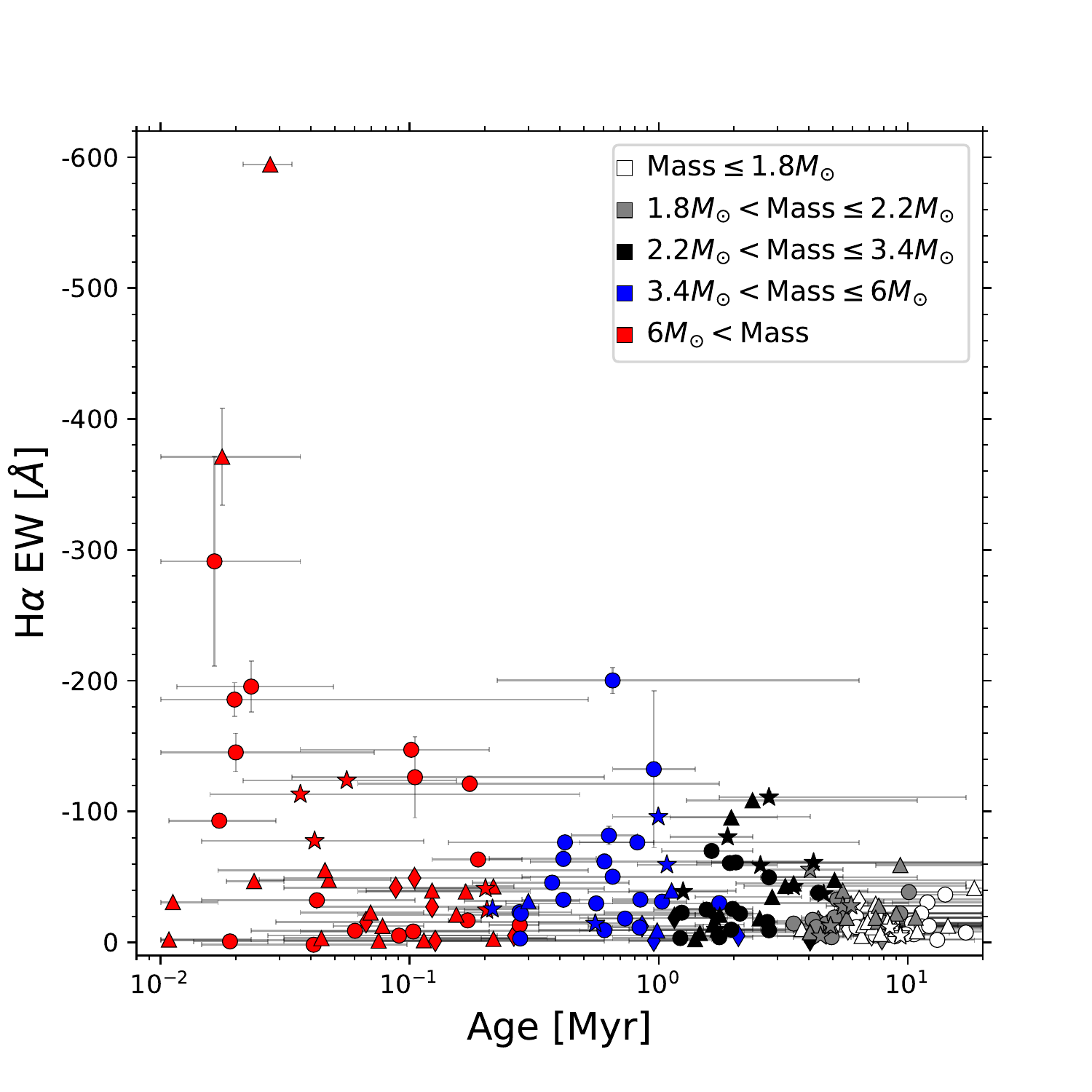}
\caption{Left: H$\alpha$ EW vs. estimated mass. Right: H$\alpha$ EW vs. estimated age. Effective temperatures and masses are respectively colour coded in the legend.  The symbols stand for the H$\alpha$ line profiles: circles (double-peaked), triangles
  (single-peaked), stars (P-Cygni profile) and diamonds (no information).} \label{ew}
\end{figure*}

Fig.~\ref{ew} shows the EW as a function of mass and age respectively.
As the definition of a Herbig Ae/Be star includes the presence of
emission, which is mostly from the H$\alpha$ line, it may not come as
a surprise that essentially all measurements are negative (\textit{i.e.} tracing
emission).

The EWs show a large range of values, which appears to increase with
increasing mass and decrease with increasing age (studied by
\citealp{Manoj}).  The older objects typically have lower EW's than
younger objects. It is tempting to read an evolutionary effect in this
finding - after all it would be expected that the accretion (and
therefore emission) would decrease when the PMS objects are closer to
the MS. However, we should recall that there is a strong correlation
between the age and the mass of the stars, so we may well be looking
at a mass effect instead.  As the EW is a relative measurement with
respect to the stellar continuum, a larger EW for otherwise similar
objects indicates a stronger emission line.  The observed trend
towards higher temperatures/masses and thus higher luminosities
implies that the lines become even stronger than the EW alone would
seem to imply.

\subsection{H$\alpha$ EW and infrared excess}\label{sec:New_IR}

\begin{figure}
  \includegraphics[scale=0.64]{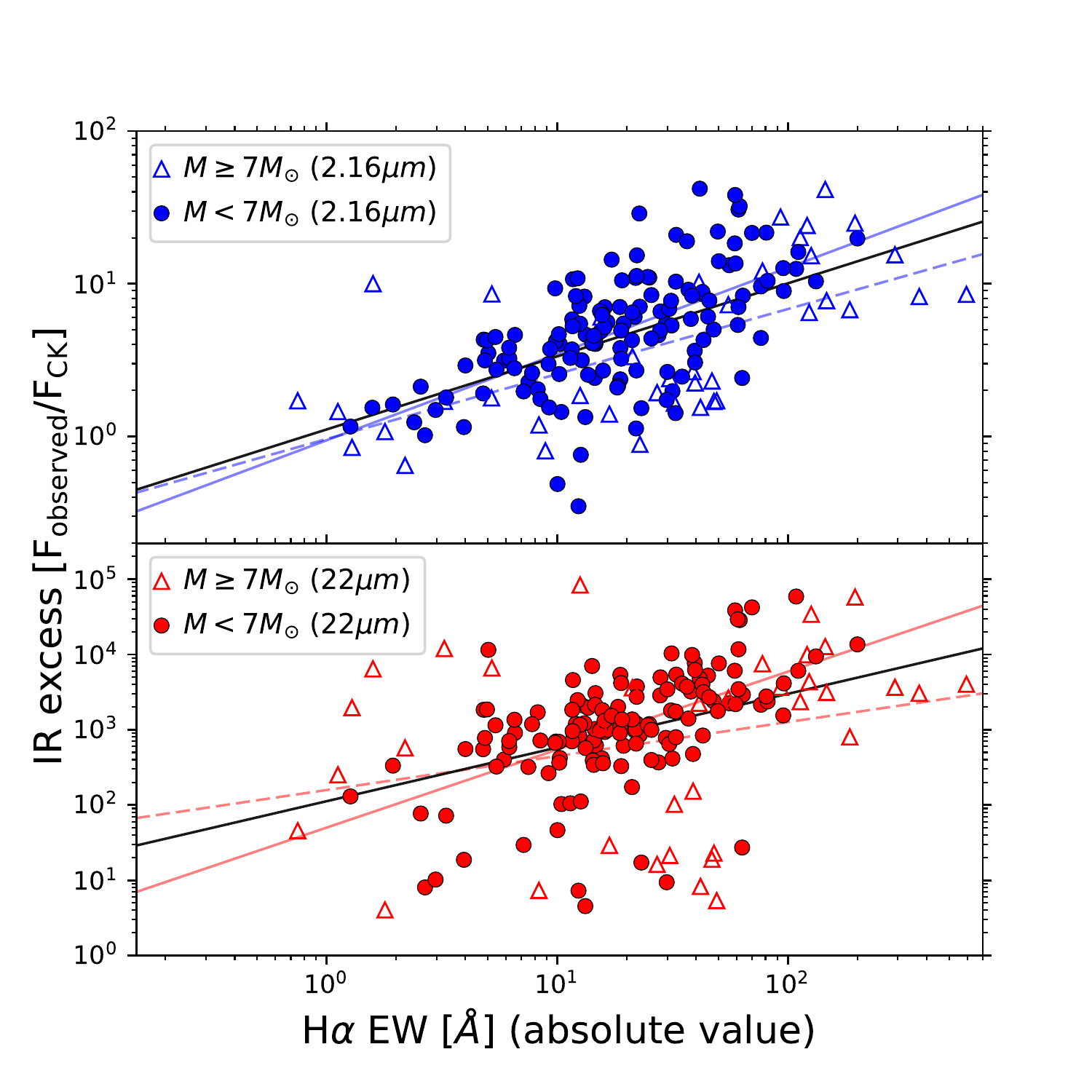}
    \caption{$2.16\mu$m (blue markers) and $22\mu$m (red markers)
    infrared excesses defined as $F_{observed}/F_{CK}$ vs. H$\alpha$
    equivalent width (absolute value). Note that this IR excess
    indicator is a flux ratio and not the one described in
    Eq.~\ref{eq1} where we integrated under the SED. Dots are Herbig Ae/Be stars with $M<7M_{\odot}$ and
    {triangles are Herbig Ae/Be stars with $M>7M_{\odot}$. Lines
    are linear fits to the data, dashed for HAeBes with $M>7M_{\odot}$
    and in solid colours for HAeBes with $M<7M_{\odot}$; black solid lines are
    the linear fits for all the sources (equations and
    correlation coefficients for these fits to all the sources for all the infrared
    bands can be seen in Table~\ref{table1}). Note the difference in the scale
    of the vertical axis between the two panels.}
  } \label{irvsew}
\end{figure}

The correlation between H$\alpha$ emission, measured by its equivalent
width, and near- and mid-infrared excess is studied in
Fig.~\ref{irvsew} and Table~\ref{table1} for each one of the infrared bands (J,
H, K\textsubscript{s}, W1, W2, W3 and W4). In this case, we computed
the infrared excess as the flux ratio between the dereddened observed
monochromatic flux and the expected flux according to the CK model at each band ({the values for these excesses
are presented in Table~\ref{table4} and Table~\ref{table4_2} for the 
high quality and low quality samples respectively}).
In all cases, there is a general and consistent increase of the H$\alpha$
EW from sources with very little IR excess to those with higher
IR excess. 


In Table~\ref{table1} we show that the H$\alpha$ emission line
equivalent width is more correlated with the infrared excess at
shorter wavelengths than at larger wavelengths, with the correlation
peaking at $2.16\mu$m (K\textsubscript{s} band). 

An obvious question might be whether there is a
causal correlation between the H$\alpha$ emission and presence of
emission due to dust around these objects. The various excesses at
various wavebands are correlated with each other (Fig.~\ref{irvsir}),
and as a consequence the IR excess at many wavelengths also correlate
with the EW. However, the correlation with H$\alpha$ is strongest at
the K\textsubscript{s} band which traces the hot dust in the inner disk, suggesting
the accretion mechanism or wind activity as traced by H$\alpha$ is
related to the inner parts of the dusty disk (see also \citealp{Manoj}). 
As presented in Table~\ref{table1},
the correlation rises from {a minimum at} $1.24\mu$m (effectively tracing the
stellar photosphere) up to $3.4\mu$m and then goes down again {to the same}
minimum at $22\mu$m (W4 band), where dust in the outer disk is found. In fact,
\citet{Mendigutia} discovered the same correlation between IR excess
and accretion rate and they found that it is no longer present beyond
$20\mu$m.

{For comparison purposes, in Fig.~\ref{irvsew} the K\textsubscript{s} band is plotted in the upper
panel and the W4 band in the lower. It is noteworthy that for the K\textsubscript{s} band, where we have the strongest correlation, small EWs are {almost} only present in sources
with {little} IR excess and, in consonance with Sect. \ref{sec:IR_excess},
for a given H$\alpha$ EW value low mass stars ($M<7M_{\odot}$) tend to have higher IR
excesses. However, these trends are weaker or non existent in the case of the W4 band, where
we have the weaker correlation. This reinforces the idea that the H$\alpha$ emission
is correlated with the inner parts of the disk. Note that in both panels the average excess is still lower for the higher mass objects.}
The emission line strengths will also be subject of a follow-on study
using accretion rates (Wichittanakom et al. in preparation).


\subsection{Variability}\label{Var_analysis}

We conclude this section by studying the variability of the objects
and its correlation with the various properties discussed so far,
including the H$\alpha$ line profiles we took from the literature.

The left panel of Fig.~\ref{ewvsire} presents the variability indicator as function of
the total (near plus mid) IR excess. As described in
Sect. \ref{sec:Variability} the variability indicator states the
number of standard deviations a certain source is separated from the
mean of the Gaia objects of the same brightness.  No, or hardly any
variability is present at the lowest IR excesses but sources can be
both variable and non-variable at the higher IR excesses, consistent
with \citet{van den Ancker3} based on a smaller sample.

The figure next to it shows the variability as a function of mass. As
high mass stars in this sample generally do not have a strong IR
excess, we find that {mostly} the lower mass and cooler objects display
{high variabilities}, with the break also around 7
M$_{\odot}$, corresponding to a Main Sequence spectral type of around
B3. {Although cooler objects tend to have larger variabilities (also observed
by \citealp{van den Ancker3}), we can observe how the range in temperatures 
for variable sources is wide in the right panel of Fig.~\ref{ewvsire},
and that there are in fact many Herbig Be stars with very strong variabilities. Hence, this is more likely a trend with mass and not with temperature. We do note that although we detect photometric variability from the $V_{i}=2$ value, the $V_{i}=5$ value is a better separation boundary for the observed trends in both panels of Fig.~\ref{ewvsire}.}

The challenge is to identify which property lies at the cause of the
variability, is it the mass of the objects, their age or infrared
excess emission or something else?  An important clue is that
many objects with strong variability (above $V_{i}=2$) {and line shape information}
have doubly peaked H$\alpha$ profiles ( {31 out of 43; $72\pm$7\%, 68\%
confidence interval)}. In general, double-peaked emission line profiles
are due to rotating disks, so the data are suggestive of an edge-on
disk-type orientation and structure {(from the remaining  {12} objects they all
have a P-Cygni profile and none have a single-peaked profile)}. The
number of variable objects with doubly peaked line profiles is
significantly different from the full sample, in which only half of
the targets with known line classifications have a double-peaked
profile {(of the sources with derived variability indicator and known line profile
 79 out of 155; 51$\pm$4\% are double-peaked and 48 our of 155; 31$\pm$4\% are single-peaked)}. 
 These fractions are significantly
different, and we therefore suspect that the variable sources are
mostly oriented edge-on, and that the line-of-sight inclination to the
objects could be a decisive factor in the cause of the variability.
{This is in agreement with the trend observed in the left panel of Fig.~\ref{ewvsire}. Sources with large amounts of circumstellar
material show large infrared excesses and high or low levels of variability depending on the inclination of their disk whilst sources with little material around
have low infrared excesses and low variabilities in all cases (also discussed in \citealp{van den Ancker3}).}




\section{Discussion}\label{sec:Discussion}

\subsection{General findings}

In the above we have determined fundamental parameters such as
temperature, mass, age, IR excess, variability and luminosity for a
large sample of Herbig Ae/Be stars which was made possible due to the
more than a factor of ten increase in available distances to these
objects compared to Hipparcos. With the {Gaia DR2} data, the majority of
known Herbig Ae/Be stars could be placed in the HR diagram. We found
the following:

\begin{itemize}

\item There are more low mass objects than high mass objects, with the
  high mass objects mostly located close to the Main Sequence.


\item High mass objects have {in general} very small infrared excesses and low
  variability, the properties appear to differ around 7M$_{\odot}$.
  
\item H$\alpha$ emission is generally correlated with infrared excess,
  with the correlation stronger for IR emission at wavelengths tracing
  the hot dust closest to the star.
  
\item More massive and younger objects have higher H$\alpha$ EWs.  

\item When split at 7M$_{\odot}$ into ``low'' and ``high'' mass
  samples, the H$\alpha$ - IR excess correlations hold for both mass
  ranges, with the average excess lower for the higher mass objects.
  
  

\item Photometric variability can be traced back to those objects with double-peaked H$\alpha$ emission
 {and large infrared excesses.}
  
\item All catalogued UXORs in the sample with detected
  variabilities above $0.5$mag in the V band appear as strongly variable (above $V_{i}=2$) with the exception of BO Cep (discussed in Sect. \ref{sec:Variability}).
  
\end{itemize}

Here, we will discuss these findings and their implications for the formation of
intermediate mass stars. 

\subsection{Selection effects}\label{sel_effect}

Let us first investigate the various
selection effects and biases that could potentially affect the
results.

\medskip

\noindent
{\it Quality parallaxes:} It could be argued that the quality of the
astrometric data has an effect on the findings. The parallax errors
occupy a comparatively small range, from
{$\sim$0.016-0.37~mas}, but because of the large spread in
distances, the relative uncertainties can be very large. To
investigate whether this has a detrimental effect on the results, we
repeated the analysis with only the objects with the very best
parallaxes {($\varpi / \sigma_{\varpi} > 10$)}. This, of
course, limits the sample and {182/218} objects remain 
{in the high quality sample}. These
{182} objects are less luminous, which may be expected as in
general they have larger parallaxes and are thus closer. As a result
they will be less massive and have larger ages than the objects in the
entire sample. {This, as a consequence of the trends described
  in previous sections, implies that these objects also show larger
  infrared excesses and variabilities as well as smaller H$\alpha$ EWs
  (see Fig.~\ref{irvsm}, Fig.~\ref{ew} and Fig.~\ref{ewvsire})}.
However, we find that essentially all correlations also hold for the
higher quality parallax sample, {and if anything, they appear
  stronger.  For example almost all of the high mass sources that 
  have large infrared excesses and variabilities in figures Fig.~\ref{irvsm} and
  Fig.~\ref{ewvsire} have $\varpi / \sigma_{\varpi} < 10$.}
  The inclusion of lower quality parallaxes induces an extra scatter in
  the results, but the larger sample and wider coverage in luminosity
  aids in reinforcing them.

\medskip

\noindent
{\it Quality identification as Herbig Ae/Be star:} Another potential
source of error is source misclassification. We have used the largest
sample of Herbig Ae/Be stars published to date (\citealp{Chen} with
some added from {{\citet{Alecian}; \citet{Baines}; \citealp{Carmona}; \citealp{Fairlamb}; 
\citet{Hernandez2}; \citet{Manoj} and \citet{Sartori}}}). The
defining characteristics of HAeBes are not unique to the class,
and can often also be found in other types of stars such as classical
Be stars, which display H$\alpha$ emission and a near-infrared excess
(\textit{e.g.} \citealp{Rivinius}) or evolved stars which can have
spectral types A and B, display hydrogen recombination emission and be
surrounded by dusty shells and disks such as the Luminous Blue
Variables and B[e] stars (\citealp{Davies}; \citealp{Oudmaijer7} on \object{HD
87643}). It is therefore inevitable that some sources will have been
misclassified. It would be fair to say that the more ``classical''
Herbig Ae/Be stars going back to the \citet{Herbig} and \citet{The} papers
have been studied in more detail and are better established as young
pre-Main Sequence stars.

We therefore studied the The et al. sample of objects (their Table 1,
{85 sources out of our 218}) separately and find that all correlations
do hold for this ``gold standard'' sample as well. We do find that on
average these objects have a larger H$\alpha$ EW
and have larger IR excesses than the full sample. These properties are
the defining characteristics of a Herbig Ae/Be star, and it may not be
surprising that the first  objects to be proposed as Herbig Ae/Be
stars are on average more extreme in these properties. Yet, again, as
with the higher quality parallax sample, the trends are still present
in this sub-sample.

\begin{table}[t]
\label{table1}
\centering
\caption{Correlation between IR excess and H$\alpha$ EW at different wavelengths.}
\begin{tabular}{lrrr}
\hline\hline 
Band & Correlation & A & B\\
 & coefficient (r) & &\\
\hline
J ($1.24\mu$m) & $0.41$ & $0.15\pm{0.03}$ & $0.025\pm{0.034}$\\
H ($1.66\mu$m) & $0.56$ & $0.32\pm{0.03}$ & $0.0024\pm{0.0478}$\\
{K\textsubscript{s}} (\boldmath$2.16\mu$m) & \boldmath$0.60$ & \boldmath$0.48\pm{0.05}$ & \boldmath$0.046\pm{0.066}$\\
W1 ($3.4\mu$m) & $0.57$ & $0.64\pm{0.07}$ & $0.15\pm{0.10}$\\
W2 ($4.6\mu$m) & $0.57$ & $0.78\pm{0.09}$ & $0.24\pm{0.12}$\\
W3 ($12\mu$m) & $0.52$ & $0.93\pm{0.12}$ & $0.79\pm{0.16}$\\
{W4} (\boldmath$22\mu$m) & \boldmath$0.41$ & \boldmath$0.71\pm{0.12}$ & \boldmath$2.05\pm{0.17}$\label{table1}
\\
\hline
\end{tabular}
\tablefoot{Correlation between IR excess (defined as a flux ratio, F\textsubscript{observed}/F\textsubscript{CK}) and H$\alpha$ EW at different wavelengths for all the sources. The coefficients are defined by:
$\log($F\textsubscript{observed}/F\textsubscript{CK}$)=A\log(|\rm EW|)+B$. {The K\textsubscript{s} band,
with the higher correlation, and the W4 band are in bold;} both are shown in Fig.~\ref{irvsew}.}
\end{table}

\medskip
\noindent
{\it Mass distribution of the sample:} The known Herbig Ae/Be stars
have mostly been found serendipitously, and a large-scale systematic
search for them has yet to be carried out. Yet, an interesting
question is how representative the present sample is for the class. To
this end, we consider the mass distribution of the objects. There are
more or less the same number of low mass, A-type objects than higher
mass B-type objects. There are more Herbig Be stars than might be
expected from the Initial Mass Function, however, the B-type objects
are brighter and are sampled from a larger volume, as also attested by
their smaller parallaxes. We will therefore expect a larger fraction
of Herbig Be stars in the sample. When limiting our sample in
distance, we obtain a Herbig Ae/Herbig Be ratio that is close to the
IMF. As far as the mass distribution is concerned, we may say that the
current sample is representative of the class. One of our future goals
is to draw an increased and well-selected sampled of Herbig Ae/Be
stars from the Gaia catalogues.

\medskip
\noindent
{\it Binarity:} One may think that binarity may affect the observed
photometry and for example produce fake levels of variability in our
variability indicator. This is because binary sources tend to be more
astrometrically and photometrically irregular. We studied the group of
binaries against the group of isolated sources and overall we find
that the known binaries are slightly brighter than the objects that
have not been reported to be a binary. This is probably a selection
effect in that brighter objects were more likely to be included in the
binary surveys. We compared the brightnesses of binaries and
non-binaries in the \citet{Baines}, \citet{Wheelwright} and
\citet{Leinert} studies separately and find that
within the surveys there are indeed no brightness differences between
binaries and non-binaries.

Returning to the Gaia sample; all other properties {but infrared excesses}, including
variability, are similar.  {We do find that binaries have in general slightly larger IR excesses}.
With the benefit of hindsight, this is
perhaps something that could have been expected. Most of the
binaries are distant binaries with separations larger than 0.1 arcsec {(Gaia's resolution)}.
Indeed, no binaries are found closer than 30au so it is expected that
binarity does not play a significant role in the optical
photometry. {At the same time}, companions could potentially contribute to the infrared
emission whose fluxes have been measured with apertures larger than
the typical separations. Given that we {do detect slight differences
in IR-excess between binaries and non-binaries, a preliminary
inference would be that the companions may contribute
to the infrared flux in some cases.}

\bigskip

\subsection{Infrared excess as function of mass}\label{sec:IR_discussion}

Fig.~\ref{irvsm} shows the infrared excess as function of mass and age
respectively.  There is a marked difference in the infrared excess
observed towards high and low mass objects. Herbig Be stars more
massive than $\sim$7 M$_{\odot}$ {in general} appear to have little to no excess,
while the lower mass objects show a wide range of excesses.  There is
also a trend with age with the youngest objects having the smallest
infrared excess. Although it would be tempting to assume a causal
relation between age and presence of dust, and try to explain why the
youngest objects have the smallest amount of dust around them, we
suspect the stellar mass is the dominant factor. The durations of the
PMS evolutionary tracks are progressively shorter for higher masses,
and an underlying relation between mass and infrared excess would
therefore also appear as a correlation between age and infrared
excess.

Either way, the lack of dusty emission from high mass objects is
puzzling, as we might expect the more massive objects to be formed in
more massive clouds and therefore be more embedded. A natural
conclusion would be that at any time of their PMS evolution, these
young objects would be surrounded by more dust than their lower mass
counterparts and therefore, at any stage, they would have a stronger
infrared emission.  A counterargument is that the Herbig Be stars are
predominately found closer to the ZAMS and are therefore more evolved,
having dispersed their circumstellar material. Supporting this idea,
\citet{Alonso-Albi} found, from their compilation of millimetre
observations of 44 objects, that Herbig Be stars have much weaker
millimetre emission than their later type counterparts.
In addition, they found that the masses of the disks around Herbig Be stars traced
at mm wavelengths are usually 5-10 times lower than
those around lower mass stars, with the boundary also around 7
M$_{\odot}$. These authors suggest that the disk dispersal is more
efficient and faster in high mass objects above 7~M$_{\odot}$.
Indeed, the disk dispersal times are a steep, declining function with
stellar mass, from millions of years for the lower mass stars  to
tens of thousands of years for the highest mass young stars of 10
M$_{\odot}$ and higher (\citealp{Gorti}).

The latter timescales are comparable to the evolutionary timescales as
for example computed by \citet{Bressan} for these massive objects.
Thus, the observation here is consistent with the classical scenario
that the Kelvin-Helmholtz contraction timescale is much smaller for
massive objects compared to the free-fall timescale of the collapsing
parental cloud. In this scenario, the massive young stars only become
visible once they are on, or close to, the Main Sequence - the
so-called birthline. We will discuss more on this later, but
{note that with this interpretation one still would expect a
  range of infrared excesses in any sample. This is consistent
  with what we find for massive objects (larger
  than 7~M$_{\odot}$); a large number of low excess stars, but still 
  a few with noticeable excess (see Fig.~\ref{irvsm}).}

Moving to the lower mass objects, which do display a large range of
infrared excess emission, an immediate question to ask is whether we
can detect any evolutionary effect in the sense that objects that are
further evolved have smaller infrared excesses, as one expected from
the progressive dust dispersal, and as suggested by \citet{Fuente}.
For example, if the inside out clearing model of disk evolution is
correct, we should see a trend at each PMS track from high excess to
little excess.

However, it appears Herbig Ae/Be stars do not show any consistent
evolution of the infrared excess from high to low excess at any
mass range.  There are many objects appearing younger than 2.5~Myr or
even 1~Myr at all mass ranges with little IR excess.  Arguably the
lack of an evolutionary effect can be explained by the size of the
error bars on for example the luminosity. The evolutionary timescales
vary strongly with mass (and thus luminosity), masking any trend of
infrared excess emission with age.  Here, we would highlight that many
young Herbig Ae stars show little excess. By looking at these objects in the right panel of Fig.~\ref{hrd2} 
it is not difficult to find sources with error bars
small enough to discard the contribution of uncertainty to the problem.
{Finally, the contamination
by binaries as discussed in Sect. \ref{sel_effect} can play a role in here
as many HAeBes can still remain as undetected binaries.}

We should also note that the underlying assumption of the evolutionary
calculations is that the conditions under which the stars form are
uniform, the accretion rates a smooth function of time resulting in an overall
similar evolution for all stars. However, the final configuration is
undoubtedly affected by inhomogeneities, varying accretion rates and
even the masses of the initial clouds.  Nevertheless, looking for real
evolutionary effects in the spectral energy distributions requires
selecting subsamples of objects that are located at or close to the
same mass tracks. This may require even more precise parallaxes
than can be provided by Gaia at the moment {in many cases. It
also requires precise determinations of the atmospheric parameters and extinction
values. A proper statistical study with high quality parameters of the evolutionary properties
of the HAeBes as they move towards the Main Sequence is hence still pending and
planned for the future.}

\subsection{Variability in terms of the UXOR phenomenon}\label{sec:Var_discussion}

\begin{figure*}[ht!]
  \includegraphics[scale=0.61]{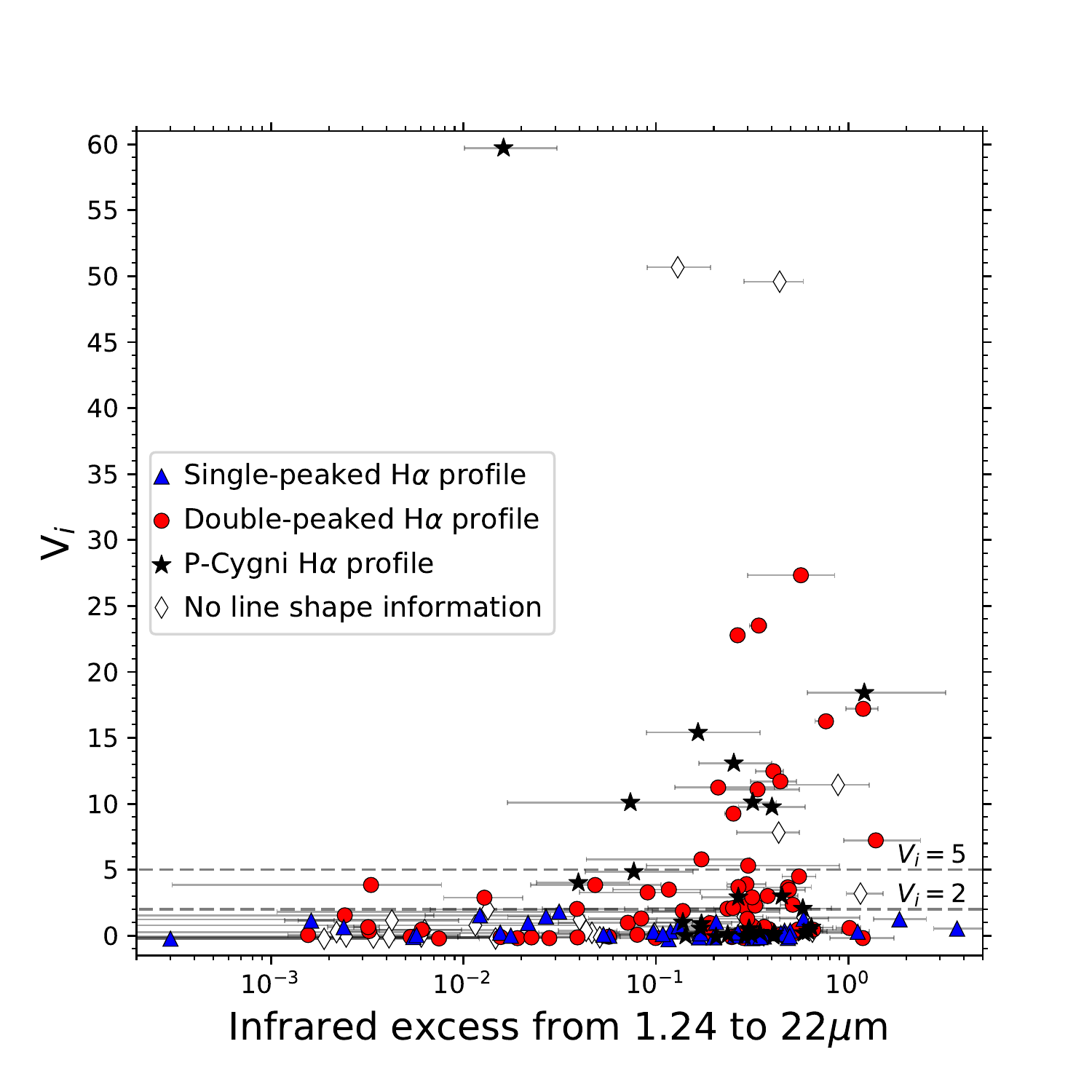}
  \includegraphics[scale=0.61]{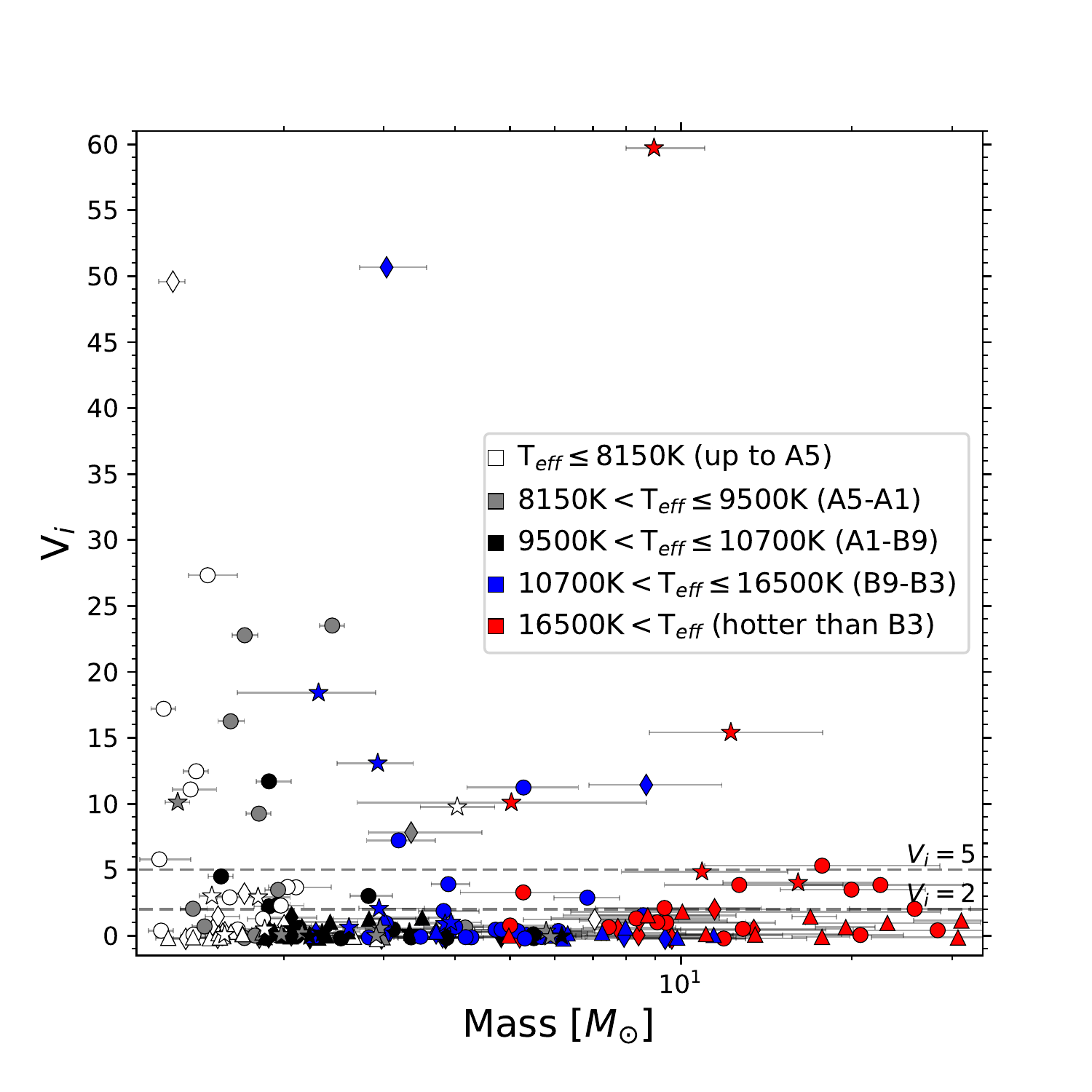}
  \caption{Left: Variability indicator vs. IR excess in the range $1.24-22\mu$m. It can
  be seen how objects with the lower IR excess do not show high variability. Right: Variability indicator vs. estimated mass. It can
  be seen how the most massive objects (more massive than $\sim7
  M_{\odot}$) do barely show variability. Line profiles and temperatures are colour coded in the legend in the left and right panel respectively.  The symbols stand for
  the H$\alpha$ line profiles: circles (double-peaked), triangles
  (single-peaked), stars (P-Cygni profile) and diamonds (no information). The $V_{i}=2$ and $V_{i}=5$ values are stressed for clarity.}\label{ewvsire}
\end{figure*}

The variability indicator that was developed specifically for the Gaia
data demonstrates that the class of Herbig Ae/Be stars is more
variable than the general population of stars.  Fig.~\ref{ewvsire}
shows that the lower mass objects are much more photometrically
variable than the higher mass objects, for which the variability
appears to cease beyond $\sim$7~M$_{\odot}$. The photometrically
variable objects contain most of the so-called UXOR variables reported
in the literature.
Using the compilation of UXOR variables by \citet{Mendigutia_thesis}; \citet{Oudmaijer5} and \citet{Poxon}, 
we find that {17 out of the 48} strongly variable objects - those with variability
indicator values larger than 2, representing variations of 0.5
magnitudes (in the V band) or higher - are classified as UXORs.
The remaining 5 UXORs {with variability indicator values} present in the sample have documented variabilities
below 0.5 magnitudes {with the exception of BO Cep (discussed in Sect. \ref{sec:Variability})}.  


The defining characteristic of the UXOR phenomenon is not only the
photometric variability but also the reddening and blueing associated
during the variations.  The explanation put forward for this behaviour
is the obscuration of the star by a rotating, inhomogeneous, dusty
edge-on disk. The objects first become redder when dust obscures the
object, and can even become blue at their faintest phases, when the
direct light from the stars is blocked and, predominately blue light
is scattered into the line of sight. As the polarization - resulting
from scattered light - also peaks during the faintest phases
(\textit{e.g.} \citealt{Grinin}), the obscuring disk hypothesis is
favoured. Interestingly, observational evidence other than the
polarization supporting this conclusion has been relatively sparse.


With the large sample of Herbig Ae/Be stars, and the large number of
UXORs among them, we can repeat a similar experiment using the
H$\alpha$ line as a proxy for the inclination of the circumstellar
disks.  We will consider the line profiles of the H$\alpha$ emission
in tandem with the variability indicator. Fig.~\ref{ewvsire} shows
that all but {twelve} of the strongly variable objects 
{with documented line profiles} (above $V_{i}=2$, those with
$\Delta V>0.5$mag) have double-peaked H$\alpha$ emission. In fact,
the {five} objects for which no line profile is listed have, to
our knowledge, no reported profiles. The occurrence of double-peaked
profiles in the highly variable sample is significantly higher than for the
other objects (see Sec.~\ref{Var_analysis}). {It is significant that 
the other twelve objects have P-Cygni profiles and none of them
show a single-peaked profile. The P-Cygni profile is often related
to episodic energetic phenomena and it is not unexpected that it is also traced by 
our variability indicator}. 
Given that doubly peaked line
profiles are most easily explained by at least part of the emission
originating in a rotating disk leads us to conclude that the
photometrically variable objects are seen edge-on and surrounded by a
disk-like structure. {It is true that outflows or winds not limited to
the disk can produce double-peaked H$\alpha$ profiles (\citealp{Kurosawa2}; \citealp{Tambovtseva}). 
Supporting the hypothesis of edge-on disks being the main cause of photometric variability, we find in variability the same 
separation at $\sim$7~M$_{\odot}$ between low and high mass objects 
we found when studying IR excesses, which suggests that photometric variability and IR excess have the same cause. In addition, sources with high IR excesses have both high and low variability levels, 
which can be understood as depending on the disk inclination, while sources with lower IR excesses show little variability in all cases (left panel of Fig.~\ref{ewvsire}, discussed in Sect. \ref{Var_analysis}). 
This would also explain the few high mass strongly variable objects 
that can be seen in the right panel of Fig.~\ref{ewvsire}; they are mostly the ones with
high IR excess in the left panel of Fig.~\ref{irvsm} (discussed before in Sect. \ref{sec:IR_discussion})}. 
Given that an edge-on orientation is the major
and main ingredient of the dust obscuration hypothesis, these results
lend very strong support to it using a large sample of Herbig Ae/Be
stars.

The large fraction of objects with double-peaked line profiles
or variability  is in agreement with the
model predictions by \citet{Natta2} who worked out how many Herbig Ae/Be
stars would undergo the UXOR phenomenon considering the scale heights
of dusty disks and under which inclinations the photometric
variability would still be visible. They conclude that around half of
the Herbig Ae stars could be UXORs.
{In our high quality sample we have $85$ A type stars with variability indicator values
and just $16$ of them were previously listed as UXORs, but again
most of them have been largely unstudied. However, of the $25$ A type
stars with variabilities above $V_{i}=2$, $13$ are known UXORs. This means that
for the Herbig Ae stars for which we detect variability at the $V_{i}=2$ level, $\sim52\%$ are known UXORs (and
just two have P-Cygni profiles).
Moreover, this implies that we are retrieving $\sim81\%$ of known A-type UXORs
with our variability indicator and hence, assuming that all the $25$ A type
stars with variabilities above $V_{i}=2$ are of UXOR type, there should be about
31/85 UXORs in the sample. In turn this would imply that $\sim37\%$ of all Herbig Ae
stars belong to the UXOR class. 
If we also take into account that we
have potentially removed some UXORs, possibly the most variable ones, from consideration when applying
the constraints described in Sec.~\ref{sec:Variability} we get to values close
to the $50\%$ predicted by \citet{Natta2}.} 

{Finally, \citet{Davies3} recently studied in detail the UXOR object \object{CO Ori},
which has single-peaked H$\alpha$ emission. Consequently, they found that the inclination of its 
disk is of $\sim 30^\circ$ (\textit{i.e.} it is nearly face-on). 
In this particular case, whether if the disk is still causing the UXOR phenomenon or if it is caused through fluctuations in the circumstellar
material outside the disk is still uncertain. We could not derive a variability indicator value for this object to assess its variability. Inspired by this example, 
we took a look to the other UXORs 
in our sample with single-peaked profiles, they all 
have variabilities below $V_{i}=2$ in our variability indicator (\object{HD 100546}, \object{HD 142527}, \object{HD 98922} and \object{IL Cep}), this 
suggesting a category of low variability UXORs with nearly face-on disks. Nonetheless, the
results presented in this section strongly support the idea that most UXORs are caused by edge-on disks, which are responsible of large photometric variabilities.}




\subsection{Missing objects in the HR diagram}

When inspecting {the right panel of Fig.~\ref{hrd2}}, it appears that most Herbig Be stars
are located relatively close to the Main Sequence, whereas the lower
mass Herbig Ae stars occupy a larger part of their evolutionary
tracks, contracting to higher temperatures at constant luminosity. In
other words, the late type Herbig Be and Herbig Ae stars at high
luminosities (and low surface gravities) that would occupy the tracks
towards the locations of B-type stars on the Main Sequence are
missing. It is only due to the use of Gaia parallaxes, expanding the
number of Herbig Ae/Be stars with well established luminosities that
we can make this observation.

In our discussion earlier, we mentioned the fact that these objects
could still be heavily embedded in their parental clouds, preventing
them from being optically visible when evolving on their way to the
Main Sequence. There is evidence for optically invisible, but infrared
bright objects at locations in these regions of the HR-diagram. For
example, \citet{Pomohaci} were the first to spectrally type an
infrared bright Massive Young Stellar Object based on the, rare,
absorption spectrum at near-infrared wavelengths (higher order
Brackett lines are in absorption for this object, while Br$\gamma$ is
in emission). They found that the object could be fitted with that of
an A-type giant star. Had this object been optically visible, it would
have occupied the empty region in {the right panel of Fig.\ref{hrd2}}. To this, we add the
early B-type Herbig Be stars/infrared bright MYSOs \object{PDS 27} and
\object{PDS 37} (\citealp{Ababakr3}).  They are found in the upper
regions of the HR diagram, slightly off the Main Sequence. They are
optically visible, but not overly bright at $V\sim13$mag, and have not
been included in many (optical) magnitude limited catalogues. 
Thus, there are several examples that might lead us to conclude
that the - implicit - optical brightness limit of any 
catalogue of Herbig Ae/Be stars would prevent the inclusion of
massive pre-Main Sequence stars on the horizontal portions of the
evolutionary PMS tracks. {However, these object are present
in Gaia DR2 although they are yet uncatalogued as HAeBes}. 
In a sense this is a situation similar to
that outlined for the low infrared excesses observed toward the Herbig
Be stars that are mostly located close to the Main Sequence.  This
could be explained by the fact that the objects would be embedded and
thus optically invisible or faint in earlier phases of their
evolution.

Further observations of optically fainter objects will be necessary to
settle this issue. Additional progress can be made by connecting the
PMS evolutionary tracks with radiative transfer codes to provide
synthetic observations (as \textit{e.g.} \citealp{Davies2}, or \citealp{Zhang2}
for Massive Young Stellar Objects) extended to optical wavelengths in
the Herbig Be mass range. Related to the ``missing'' high mass stars
in the HR diagram, it will be important to fill the historic, and
entirely man-made, gap between the Herbig Ae stars and the T-Tauri
stars. The latter are confined to have spectral types G-K-M, and
typically Herbig Ae/Be stars, in this case by definition, have
spectral types A and B. We are missing out the F-type stars, resulting
in an incomplete coverage of the HR diagram for pre-Main Sequence
stars.

\subsection{On the difference between Herbig Ae and Herbig Be stars}

From the above it appears that the dusty disks surrounding Herbig Ae
and Herbig Be stars are different, with the break in IR excess
occurring at 7M$_{\odot}$ (around B3 spectral type), a value which was
also found by \citet{Alonso-Albi} from their compilation of millimetre
emission tracing the outer parts of the dusty disks. As discussed,
given the much stronger radiation field from B-stars, both in
intensity and photon-energies, the most straightforward explanation
for the much less massive disks of higher mass objects is a more
efficient disk dispersal mechanism (see \textit{e.g.}
\citealp{Gorti}).  This also explains why the same 7M$_{\odot}$ break
is seen in variability (Fig.~\ref{ewvsire}).  As described in
Sect. \ref{sec:Var_discussion}, the high levels of variability in some
sources are caused by edge-on dusty disks. A more efficient disk
dispersal mechanism beyond 7M$_{\odot}$ would result in these sources
showing no strong variability in our indicator. It also explains
why the objects with the lower IR excesses are not strongly
variable while the rest can have both high and low variability values.

Other studies of large samples of Herbig Ae/Be stars indicate a break
in properties at a much lower mass of 3M$_{\odot}$, around the B7
spectral boundary. \citet{Fairlamb} studied the accretion rates, which
are proportional to the mass of the objects, and found a different
slope for lower mass than for higher mass objects. \citet{Ababakr},
extending the work of \citet{Mottram}, found a distinct difference in
spectro-polarimetric properties across the H$\alpha$ line between the
Herbig Ae and late Be type stars on the one hand and earlier Herbig
Be type objects on the other hand. These authors also point out the
similarity in the H$\alpha$ spectro-polarimetry of the Herbig Ae stars
and T-Tauri stars.  Finally, \citet{Mendigutia3} noted the difference
in H$\alpha$ variability; Herbig Ae and late Be stars are largely
variable, whereas Herbig Be stars are not. Later, \citet{Fang} showed
that T-Tauri stars display even more variable H$\alpha$ emission -
again hinting at a similar accretion mechanism for the T-Tauri stars
and Herbig Ae stars.

How can we reconcile the fact that some studies show a different break
in properties than others? It is worth pointing out that the latter
investigations consider regions much closer to the star than the dusty
emission. \citet{Fairlamb} derives accretion rates from the UV-excess,
which trace the shocked material on the stellar surface,
\citet{Ababakr}'s spectropolarimetry traces the free electrons in
ionized material at distances of order stellar radii from the
stars. The spectro-polarimetric properties of the B-type stars can be
explained by stable circumstellar disks, while the line properties for
T-Tauri and Herbig Ae objects are consistent with magnetically
controlled accretion. Likewise, the H$\alpha$ emission traces the
ionized zones close to the star, such as the accretion columns and
circumstellar disks and the variability is explained due to the
accretion columns orbiting the central star (\textit{e.g}
\citealp{Kurosawa}).


Earlier, we showed that the IR fluxes and H$\alpha$ properties are
largely correlated, but that the IR fluxes are smaller for the earlier
type objects. We therefore conclude this section with the observation
that the infrared and millimetre emission trace the circumstellar
disks and originates much further from the stars than the UV, hydrogen
recombination emission and free electrons which trace the accretion
onto the stars.  The break in accretion mechanism appears to occur
around 3M$_{\odot}$, whereas the disk dispersal becomes significant at
higher masses, 7M$_{\odot}$.

\section{Conclusions}\label{sec:Conclusions}

In this paper we have collated the largest astrometric dataset of
Herbig Ae/Be stars. We present parallaxes for the vast majority of
known Herbig Ae/Be stars and have gathered atmospheric parameters,
optical and infrared photometry, extinction values, H$\alpha$ emission
line information, binary statistics, and devised an objective measure
for the photometric variability.  From these we derive luminosities
which allow us to place the objects in an HR diagram, containing over
ten times more objects than previously possible.

\bigskip

Thus, we homogeneously derived luminosities, {distances}, masses, ages,
variabilities and infrared excesses for the most complete sample of
Herbig Ae/Be stars to date.  We investigated the various properties
and reach the following conclusions:

\begin{enumerate}

\item The Gaia photometric variability indicator as developed here
  indicates that {48/193} or $\sim$25\% of all Herbig Ae/Be stars are
  strongly variable. We find that the presence of
  variability correlates very well with the H$\alpha$ line
  profile. The variable objects display doubly peaked profiles,
  indicating an edge-on disk. It had been suggested that this
  variability is in most cases due to asymmetric dusty disk structures
  seen edge-on. The observation
  here is the most compelling confirmation of this hypothesis.
  {Most sources catalogued as UXORs in the sample appear as strongly variable
  with double-peaked profiles. The fraction of
  strongly variable A-type objects is close to that found
  for the A-type objects with the UXOR phenomenon.} 
  
\item High mass stars do not display an infrared excess and show no
  strong photometric variability.  Several suggestions have been put forward
  to explain this. These include fast evolutionary timescales and fast
  dust dispersion timescales for high mass objects. We do note that
  the break is around 7~M$_{\odot}$, which is intriguingly similar to
  other statistical studies related to dusty disks around Herbig Ae/Be
  stars which signpost a different or more efficient disk dispersal
  mechanism for high mass objects.

\item Whereas the break in IR properties {and photometric variabilities} occurs at 7~M$_{\odot}$,
  various H$\alpha$ line properties including mass accretion rates,
  spectropolarimetric properties and emission line variability seem to
  differ at a lower mass, 3~M$_{\odot}$. The latter has been linked to
  different accretion mechanisms at work; magnetospheric accretion for
  the A-type objects and another mechanism, possibly boundary layer
  accretion, for the B-type objects. The differing IR and variability properties are
  related to different or differently acting (dust-)disk dispersal
  mechanisms, which occurs at much larger size scales than the accretion
  traced by hydrogen recombination line emission.

\end{enumerate}

Finally, the findings presented in this paper signal just the
beginning in unveiling the formation of intermediate mass stars using
Gaia. Gaia presents us with an excellent opportunity to search and
identify new Herbig Ae/Be stars, resulting in a well-selected and
properly characterized sample. The results presented here will assist
greatly in identifying new Herbig Ae/Be objects from the more than a
billion stars with astrometric parameters in Gaia.  This is the
subject of our follow-on study, the STARRY project.

\begin{acknowledgements}

The STARRY project has received funding from
the European Union’s Horizon 2020 research and innovation programme
under MSCA ITN-EID grant agreement No 676036.

This work has made use of data from the European Space Agency (ESA)
mission {\it Gaia} (\url{https://www.cosmos.esa.int/gaia}), processed by
the {\it Gaia} Data Processing and Analysis Consortium (DPAC,
\url{https://www.cosmos.esa.int/web/gaia/dpac/consortium}). Funding
for the DPAC has been provided by national institutions, in particular
the institutions participating in the {\it Gaia} Multilateral Agreement.

This research also made use of Astropy, a community-developed core Python package for Astronomy (Astropy Collaboration, 2013), the TOPCAT tool (\citealp{Taylor}) and the VizieR catalogue access tool and the SIMBAD database, operated at CDS, Strasbourg, France.

This research was made possible through the use of the AAVSO Photometric All-Sky Survey (APASS), funded by the Robert Martin Ayers Sciences Fund. 

We thank the referee for his/her insightful comments which have improved the paper. 

\end{acknowledgements}

\onecolumn

\begin{landscape}
\renewcommand{\arraystretch}{1.2}
\begin{longtable}{lrrrrrrrrc}
\caption{\label{table2} Main parameters of each Herbig Ae/Be star belonging to the high quality sample of 218 sources.}\\
\hline\hline 
Name & RA &  DEC & Parallax & Distance &T\textsubscript{eff} &Log(L) & A\textsubscript{V} & V & Binary\\
  & (h:m:s)  & (deg:m:s) &  (mas) & (pc) & (K)  & ($L_{\odot}$) & (mag) & (mag) & \\ 
\hline
\endfirsthead
\caption{continued.}\\
\hline\hline
Name & RA &  DEC & Parallax & Distance &T\textsubscript{eff} &Log(L) & A\textsubscript{V} & V & Binary\\
  & (h:m:s)  & (deg:m:s) &  (mas) & (pc) & (K)  & ($L_{\odot}$) & (mag) & (mag) & \\ 
\hline
\endhead
\hline
\endfoot
AB Aur & 04:55:45.9 & +30:33:04 & $6.140\pm{0.057}$ & $162.9^{+2.6}_{-2.4}$ & $9500^{+750}_{-790}$ & $1.61^{+0.19}_{-0.21}$ & $0.43^{+0.28}_{-0.35}$ & 7.32 & Yes$^{1}$ \\
AK Sco & 16:54:44.8 & -36:53:19 & $7.113\pm{0.062}$ & $140.6^{+2.1}_{-2.0}$ & $6250^{+250}_{-250}$ & $0.623^{+0.028}_{-0.005}$ & $0.000^{+0.050}_{-0.000}$ & 8.90 & Yes$^{6}$ \\
AS 310 & 18:33:21.2 & -04:58:06 & $0.390\pm{0.046}$ & $2110^{+350}_{-240}$ & $24500^{+4500}_{-5000}$ & $4.17^{+0.39}_{-0.44}$ & $4.13^{+0.26}_{-0.29}$ & 12.49 & - \\
AS 470 & 21:36:14.2 & +57:21:31 & $0.137\pm{0.027}$ & $4040^{+620}_{-440}$ & $8200^{+1600}_{-800}$ & $3.01^{+0.47}_{-0.27}$ & $2.27^{+0.62}_{-0.42}$ & 12.44 & - \\
AS 477 & 21:52:34.1 & +47:13:44 & $1.290\pm{0.029}$ & $773^{+30}_{-27}$ & $10000^{+1700}_{-500}$ & $2.22^{+0.36}_{-0.12}$ & $1.19^{+0.46}_{-0.12}$ & 10.05 & Yes$^{4}$ \\
BD+30 549 & 03:29:19.8 & +31:24:57 & $3.384\pm{0.083}$ & $295^{+13}_{-11}$ & $11500^{+1500}_{-800}$ & $1.54^{+0.25}_{-0.14}$ & $1.73^{+0.24}_{-0.12}$ & 10.51 & - \\
BD+41 3731 & 20:24:15.7 & +42:18:01 & $0.987\pm{0.035}$ & $1003^{+64}_{-53}$ & $17000^{+1000}_{-1000}$ & $2.96^{+0.31}_{-0.31}$ & $1.06^{+0.50}_{-0.50}$ & 9.90 & - \\
BF Ori & 05:37:13.3 & -06:35:01 & $2.570\pm{0.053}$ & $389^{+14}_{-12}$ & $8970^{+200}_{-200}$ & $1.287^{+0.060}_{-0.051}$ & $0.330^{+0.030}_{-0.020}$ & 9.82 & - \\
\hline
\end{longtable}
\tablebib{Atmospheric parameters T\textsubscript{eff}, A\textsubscript{V} and V taken from the following sources in order of choice: \citealp{Fairlamb}; \citealp{Montesinos}; \citealp{Hernandez}; \citealp{Mendigutia}; \citealp{Carmona}; \citealp{Chen}; \citealp{Alecian}; \citealp{Sartori}; \citealp{Manoj}; \citealp{Hernandez2}; \citealp{Vieira}; APASS Data Release 9 and the SIMBAD database. If not available they were derived as described in Sect. \ref{sec:Atmospheric}. See Sect. \ref{sec:Der_cuantities} for derivation of $L$, Mass and Age. The references for binarity are: (1) \citet{Baines}; (2) \citet{Wheelwright}; (3) \citet{Leinert}; (4) \citet{Maheswar}; (5) \citet{Wheelwright2}; (6) \citet{Alecian}; (7) \citet{Hamaguchi}; (8) \citet{Dunhill}; (9) \citet{Coulson and Walther}; (10) \citet{Liu}; (11) \citet{Biller}; (12) \citet{Schuetz}; (13) \citet{Boersma}; (14) \citet{Malkov}; (15) \citet{Ferro}; (16) \citet{Kubat}; (17) \citet{Morrell}; (18) \citet{Lazareff}; (19) \citet{Mayer}; (20) \citet{Folsom}; (21) \citet{Corporon}; (22) \citet{Doering}; (23) \citet{Chelli}; (24) \citet{Miroshnichenko}; (25) \citet{Friedemann}; (26) \citet{Kraus}; (27) \citet{Torres}; (28) \citet{Aspin}; (29) \citet{Connelley}; (30) \citet{Millour}; (31) \citet{Frasca}; (32) \citet{Marston}; (33) \citet{Zhang3}.}
\end{landscape}

\newpage

\begin{landscape}
\renewcommand{\arraystretch}{1.2}
\begin{longtable}{lrrrrrrr}
\caption{\label{table3} Other parameters of each Herbig Ae/Be star belonging to the high quality sample of 218 sources.}\\
\hline\hline 
Name & Near IR excess &  Mid IR excess & H$\alpha$ EW & H$\alpha$ & $V_{i}$ & Mass & Age\\
  & ($1.24-3.4\mu$m)  & ($3.4-22\mu$m) &  (\AA) & line shape & & ($M_{\odot}$) & (Myr)\\ 
\hline
\endfirsthead
\caption{continued.}\\
\hline\hline
Name & Near IR excess &  Mid IR excess & H$\alpha$ EW & H$\alpha$ & $V_{i}$ & Mass & Age\\
  & ($1.24-3.4\mu$m)  & ($3.4-22\mu$m) &  (\AA) & line shape & & ($M_{\odot}$) & (Myr)\\ 
\hline
\endhead
\hline
\endfoot
AB Aur & $0.32^{+0.25}_{-0.13}$ & $0.27^{+0.25}_{-0.12}$ & $-55.7\pm{4.5}^{6}$ & P$^{6}$ & 0.21 & $2.15^{+0.36}_{-0.21}$ & $4.0^{+1.4}_{-1.5}$ \\
AK Sco & $0.212^{+0.035}_{-0.053}$ & $0.196^{+0.016}_{-0.024}$ & $-5.9\pm{1.1}^{1}$ & d$^{28}$ & 12.47 & $1.401^{+0.070}_{-0.070}$ & $8.4^{+1.7}_{-0.4}$ \\
AS 310 & - & - & $-8.89\pm{0.38}^{5}$ & d$^{26}$ & -0.21 & $11.9^{+4.8}_{-3.4}$ & $0.06^{+0.54}_{-0.04}$ \\
AS 470 & $0.031^{+0.054}_{-0.031}$ & $(0.95^{+0.73}_{-0.82})\cdot10^{-2}$ & $-49.2\pm{4.0}^{23}$ & - & 1.22 & $7.0^{+2.8}_{-1.8}$ & $0.10^{+0.20}_{-0.07}$ \\
AS 477 & $0.19^{+0.07}_{-0.11}$ & $0.162^{+0.056}_{-0.092}$ & $-38.8\pm{1.5}^{22}$ & P$^{22}$ & 0.24 & $3.3^{+1.1}_{-0.4}$ & $1.25^{+0.64}_{-0.73}$ \\
BD+30 549 & $0.030^{+0.013}_{-0.015}$ & $(1.63^{+0.52}_{-0.67})\cdot10^{-2}$ & - & - & 0.22 & $2.28^{+0.37}_{-0.19}$ & $5^{+15}_{-2}$ \\
BD+41 3731 & - & $(0.24^{+0.61}_{-0.24})\cdot10^{-3}$ & - & s$^{8}$ & 0.01 & $5.0^{+1.2}_{-0.7}$ & $0.5^{+3.0}_{-0.3}$ \\
BF Ori & $0.113^{+0.011}_{-0.012}$ & $0.139^{+0.012}_{-0.013}$ & $-14.7\pm{1.5}^{1}$ & d$^{3}$ & $9.25*$ & $1.807^{+0.090}_{-0.090}$ & $6.38^{+0.32}_{-0.46}$ \\
\hline
\end{longtable}
\tablebib{The H$\alpha$ line profile classification is as follows: single-peaked (s), double-peaked (d) and showing a P-Cygni profile (P), both regular or inverse. {An asterisk together with the variability indicator indicates that the source had been catalogued as UXOR type in the literature}. References for EW values and line shapes: (1) \citet{Fairlamb2}; (2) \citet{Carmona}; (3) \citet{Mendigutia3}; (4) \citet{Ababakr2}; (5) \citet{Hernandez}; (6) \citet{Baines}; (7) \citet{Wheelwright}; (8) \citet{van den Ancker3}; (9) \citet{Oudmaijer3}; (10) \citet{Kucerova}; (11) \citet{Hernandez2}; (12) \citet{Dunkin}; (13) \citet{Pogodin}; (14) \citet{Miroshnichenko3}; (15) \citet{Sartori}; (16) \citet{Polster}; (17) \citet{Manoj}; (18) \citet{Miroshnichenko4}; (19) \citet{Miroshnichenko}; (20) \citet{Borges}; (21) \citet{Vieira}; (22) \citet{Boehm}; (23) \citet{Nakano}; (24) \citet{Spezzi}; (25) \citet{Hou}; (26) \citet{Grinin2}; (27) \citet{Vieira2}; (28) \citet{Acke}; (29) \citet{Herbig2}; (30) \citet{Oudmaijer7}; (31) X-Shooter spectra, 2015, priv. comm, from ESO observing program 084.C-0952A; (32) \citet{Ababakr};
(33) \citet{Zuckerman}; (34) \citet{Oudmaijer7}; (35) \citet{Frasca}; (36) \citet{Miroshnichenko5}; (37) \citet{Miroshnichenko6}.}
\end{landscape}

\newpage

\renewcommand{\arraystretch}{1.2}
\begin{longtable}{lrrrrrrr}
\caption{\label{table4} IR excess at each bandpass (defined as F\textsubscript{observed}/F\textsubscript{CK}) for each Herbig Ae/Be star belonging to the high quality sample of 218 sources.}\\
\hline\hline 
Name & J &  H & K\textsubscript{s} & W1 & W2 & W3 & W4\\
  & $1.24\mu$m  & $1.66\mu$m & $2.16\mu$m & $3.4\mu$m & $4.6\mu$m & $12\mu$m & $22\mu$m\\ 
\hline
\endfirsthead
\caption{continued.}\\
\hline\hline
Name & J &  H & K\textsubscript{s} & W1 & W2 & W3 & W4\\
  & $1.24\mu$m  & $1.66\mu$m & $2.16\mu$m & $3.4\mu$m & $4.6\mu$m & $12\mu$m & $22\mu$m\\  
\hline
\endhead
\hline
\endfoot
AB Aur & $2.79$ & $6.01$ & $13.23$ & $28.77$ & - & $240.49$ & $2244.34$ \\
AK Sco & $1.34$ & $1.88$ & $3.14$ & $7.78$ & $11.61$ & $49.83$ & $400.83$ \\
AS 310 & $0.68$ & $0.71$ & $0.81$ & - & - & - & - \\
AS 470 & $1.21$ & $1.28$ & $1.71$ & $2.15$ & $2.77$ & $3.21$ & $5.36$ \\
AS 477 & $2.00$ & - & - & $27.30$ & $59.04$ & $104.62$ & $475.86$ \\
BD+30 549 & $1.81$ & $2.08$ & $2.35$ & $2.79$ & $3.39$ & $51.04$ & $378.04$ \\
BD+41 3731 & $0.95$ & $0.93$ & $0.92$ & $0.94$ & $1.03$ & $2.54$ & $35.38$ \\
BF Ori & $1.50$ & $2.39$ & $4.56$ & $10.69$ & $21.87$ & $147.30$ & $617.52$ \\
BH Cep & $1.23$ & $1.79$ & $3.24$ & $7.10$ & $10.97$ & $52.96$ & $585.75$ \\
BO Cep & $1.39$ & $1.76$ & $2.27$ & $4.22$ & $5.86$ & $15.29$ & $319.78$ \\
CO Ori & $1.94$ & $2.52$ & $4.28$ & $6.55$ & $13.63$ & $36.94$ & $172.89$ \\
CPM 25 & $2.82$ & $7.14$ & $19.80$ & $73.01$ & $215.11$ & $2106.36$ & $13608.54$ \\
CQ Tau & $1.06$ & $1.92$ & $4.31$ & $8.63$ & $19.75$ & $150.12$ & $1840.10$ \\
DG Cir & $8.79$ & $15.43$ & $30.63$ & $66.60$ & $174.10$ & $1164.79$ & $11702.50$ \\
GSC 1876-0892 & $1.69$ & $3.44$ & $8.57$ & $22.43$ & $59.77$ & $650.69$ & $6589.47$ \\
GSC 3975-0579 & $1.54$ & $2.90$ & $7.03$ & $21.27$ & $36.92$ & $79.38$ & $933.24$ \\
GSC 6546-3156 & $1.57$ & $1.91$ & $2.37$ & $6.11$ & $10.31$ & $717.95$ & $5359.18$ \\
GSC 8143-1225 & $1.75$ & $2.55$ & $4.77$ & $9.99$ & $15.83$ & $26.31$ & $361.41$ \\
GSC 8581-2002 & $1.96$ & $1.97$ & $2.12$ & $1.98$ & $2.03$ & $7.17$ & $76.96$ \\
GSC 8645-1401 & $1.38$ & $2.20$ & $4.18$ & $9.16$ & $15.58$ & $69.63$ & $719.30$ \\
GSC 8994-3902 & $1.14$ & $1.18$ & $1.28$ & - & - & - & - \\
HBC 217 & $1.32$ & $1.63$ & $2.42$ & $5.27$ & $8.16$ & $33.06$ & $1021.50$ \\
HBC 222 & $1.30$ & $1.64$ & $2.74$ & $6.10$ & $9.02$ & $18.50$ & $323.71$ \\
HBC 334 & $2.04$ & $2.72$ & $3.52$ & $10.86$ & $20.90$ & $788.83$ & $11504.97$ \\
HBC 442 & $1.29$ & $1.40$ & $1.91$ & $3.40$ & $5.54$ & $69.45$ & $550.58$ \\
HBC 7 & $1.22$ & $1.26$ & $1.55$ & $1.61$ & $2.34$ & $3.56$ & $8.26$ \\
HBC 705 & $1.45$ & $1.55$ & $1.94$ & $1.79$ & $2.64$ & $4.36$ & $16.23$ \\
HD 100453 & $1.34$ & $2.00$ & $4.26$ & $9.24$ & $24.07$ & $134.69$ & $1857.99$ \\
HD 100546 & $1.34$ & $2.09$ & $3.64$ & $10.62$ & $21.19$ & $490.87$ & $7717.64$ \\
HD 101412 & $1.60$ & $2.34$ & $4.88$ & $15.60$ & $44.31$ & $209.20$ & $1006.20$ \\
HD 104237 & $1.50$ & $2.40$ & $4.60$ & $8.98$ & $21.58$ & $84.07$ & $367.56$ \\
HD 114981 & $1.23$ & $1.24$ & $1.34$ & $1.41$ & $1.39$ & $1.32$ & $4.53$ \\
HD 130437 & $2.19$ & $2.56$ & $3.24$ & $4.23$ & $5.85$ & $8.52$ & $32.43$ \\
HD 132947 & $1.30$ & $1.42$ & $1.54$ & - & - & - & - \\
HD 135344 & $0.56$ & $0.47$ & $0.49$ & $0.49$ & $0.52$ & $0.56$ & $46.31$ \\
HD 135344B & $1.38$ & $2.05$ & $4.04$ & $8.55$ & $17.29$ & $16.84$ & $420.92$ \\
HD 139614 & $1.39$ & $1.76$ & $3.16$ & $6.59$ & $12.20$ & $150.84$ & $2119.75$ \\
HD 141569 & $1.29$ & $1.32$ & $1.45$ & $2.77$ & $2.69$ & $6.48$ & $102.87$ \\
HD 141926 & $1.59$ & $1.80$ & $2.31$ & $3.01$ & $4.38$ & $5.53$ & $18.99$ \\
HD 142527 & $2.45$ & $4.16$ & $8.24$ & $15.15$ & $36.02$ & $158.75$ & $1216.10$ \\
HD 142666 & $1.64$ & $2.50$ & $4.63$ & $9.18$ & $20.00$ & $170.41$ & $908.47$ \\
HD 143006 & $1.22$ & $1.50$ & $2.56$ & $7.36$ & $14.36$ & $24.41$ & $365.00$ \\
HD 144432 & $1.86$ & $2.85$ & $5.34$ & $8.29$ & $17.05$ & $146.13$ & $837.23$ \\
HD 149914 & $1.30$ & $1.33$ & $1.42$ & $1.36$ & $1.50$ & $1.06$ & $5.65$ \\
HD 150193 & $1.90$ & $3.19$ & $6.08$ & $12.60$ & $28.58$ & $194.31$ & $980.17$ \\
HD 155448 & - & $0.94$ & $0.94$ & - & - & - & - \\
HD 158643 & $0.97$ & $1.17$ & $1.80$ & $3.31$ & - & $26.38$ & $72.09$ \\
HD 163296 & $1.81$ & $3.36$ & $7.07$ & $17.49$ & $37.20$ & $202.40$ & $853.49$ \\
HD 169142 & $1.28$ & $1.72$ & $2.71$ & $2.86$ & $4.77$ & $48.62$ & $1212.98$ \\
HD 17081 & $1.08$ & $1.10$ & $1.11$ & $1.06$ & $1.27$ & $0.69$ & $1.56$ \\
HD 174571 & $0.99$ & $1.02$ & $1.08$ & $0.67$ & $0.72$ & $0.76$ & $4.03$ \\
HD 176386 & $1.73$ & $1.77$ & $2.04$ & $3.89$ & $3.54$ & - & $1703.78$ \\
HD 179218 & $1.04$ & $1.37$ & $2.54$ & $6.64$ & $16.19$ & $289.90$ & $1956.36$ \\
HD 199603 & $0.98$ & $0.95$ & $1.03$ & $0.97$ & $1.14$ & $0.61$ & $0.76$ \\
HD 200775 & $2.25$ & $3.98$ & $8.36$ & $25.21$ & - & $149.52$ & $2914.79$ \\
HD 235495 & $3.04$ & $6.46$ & $14.38$ & $36.37$ & $74.80$ & $332.15$ & $1821.12$ \\
HD 244314 & $1.70$ & $2.81$ & $5.47$ & $13.82$ & $22.29$ & $143.73$ & $778.93$ \\
HD 244604 & $1.78$ & $3.20$ & $7.17$ & $11.50$ & $23.27$ & $137.15$ & $839.31$ \\
HD 245185 & $1.91$ & $3.16$ & $6.58$ & $17.93$ & $36.27$ & $705.27$ & $4958.59$ \\
HD 249879 & $1.31$ & $1.92$ & $5.02$ & $23.39$ & $59.15$ & $474.67$ & $2392.06$ \\
HD 250550 & $3.11$ & $7.68$ & $18.39$ & $44.57$ & $97.12$ & $614.45$ & $6073.51$ \\
HD 259431 & $2.08$ & $4.10$ & $9.61$ & $27.60$ & $94.90$ & $305.32$ & $2145.16$ \\
HD 287823 & $1.29$ & $1.99$ & $4.48$ & $12.55$ & $24.20$ & $76.66$ & $1144.68$ \\
HD 288012 & $0.63$ & $0.59$ & $0.64$ & - & - & - & - \\
HD 290380 & $1.41$ & $2.12$ & $3.72$ & $7.95$ & $12.64$ & $51.05$ & $701.84$ \\
HD 290409 & $1.67$ & $2.42$ & $4.04$ & $4.53$ & $6.53$ & $219.21$ & $3060.97$ \\
HD 290500 & $2.10$ & $2.82$ & $4.14$ & $12.80$ & $20.09$ & $100.80$ & $2143.13$ \\
HD 290764 & $1.71$ & $2.95$ & $6.34$ & $15.76$ & $28.69$ & $70.06$ & $1819.14$ \\
HD 290770 & $2.15$ & $4.07$ & $9.14$ & $19.26$ & $34.92$ & $200.05$ & $1406.43$ \\
HD 305298 & $1.46$ & $1.65$ & $1.70$ & $2.63$ & $4.76$ & $271.40$ & $12010.35$ \\
HD 313571 & $1.86$ & $2.18$ & $2.65$ & $4.29$ & $4.69$ & $10.16$ & $152.52$ \\
HD 31648 & $1.60$ & $2.68$ & $5.49$ & $9.48$ & $22.53$ & $120.18$ & $611.63$ \\
HD 319896 & $1.43$ & $1.61$ & $1.98$ & $5.75$ & $9.81$ & $22.37$ & $414.03$ \\
HD 323771 & $2.54$ & $5.90$ & $13.58$ & $38.02$ & $76.64$ & $312.52$ & $2185.92$ \\
HD 34282 & $1.82$ & $3.78$ & $8.31$ & $17.85$ & $28.81$ & $104.86$ & $1196.27$ \\
HD 344261 & $1.15$ & $1.24$ & $1.24$ & $1.23$ & $1.20$ & $0.81$ & - \\
HD 34700 & $1.00$ & $1.02$ & $1.24$ & - & - & - & - \\
HD 35187 & $1.90$ & $2.74$ & $4.65$ & $7.72$ & $13.40$ & $59.77$ & $569.45$ \\
HD 35929 & $1.35$ & $1.47$ & $1.97$ & $3.06$ & $5.72$ & $12.01$ & $29.45$ \\
HD 36112 & $1.63$ & $2.71$ & $5.61$ & $14.70$ & $31.22$ & $99.40$ & $992.76$ \\
HD 36408 & $1.06$ & $1.01$ & $1.02$ & $0.79$ & $0.81$ & $0.74$ & $8.02$ \\
HD 36917 & $1.77$ & $2.20$ & $2.98$ & $8.99$ & $20.56$ & $84.71$ & $264.73$ \\
HD 36982 & $1.80$ & $2.02$ & $2.38$ & - & - & - & - \\
HD 37258 & $1.97$ & $3.27$ & $6.62$ & $15.84$ & $28.99$ & $189.17$ & $970.64$ \\
HD 37357 & $1.54$ & $2.36$ & $4.22$ & $7.84$ & $14.86$ & $102.39$ & $696.70$ \\
HD 37371 & $0.35$ & $0.36$ & $0.35$ & $0.32$ & $0.40$ & $0.77$ & $7.26$ \\
HD 37490 & $0.88$ & $1.02$ & $1.19$ & $2.35$ & $4.17$ & $3.60$ & $7.30$ \\
HD 37806 & $2.20$ & $4.90$ & $10.98$ & $22.93$ & $60.67$ & $293.60$ & $1185.58$ \\
HD 38087 & $2.01$ & $2.45$ & $2.77$ & $2.94$ & $2.84$ & $2.04$ & $275.15$ \\
HD 38120 & $1.80$ & $3.11$ & $6.10$ & $10.25$ & $20.61$ & $658.25$ & $5227.02$ \\
HD 39014 & $0.98$ & $1.15$ & $1.17$ & $1.26$ & $1.75$ & $0.82$ & $1.00$ \\
HD 41511 & $4.36$ & $8.56$ & $12.99$ & - & - & $89.68$ & $289.86$ \\
HD 45677 & $3.19$ & $7.44$ & $32.20$ & - & - & $4700.04$ & $28602.62$ \\
HD 46060 & $0.79$ & $0.75$ & $0.74$ & $0.80$ & $0.77$ & $3.47$ & $171.16$ \\
HD 50083 & $1.23$ & $1.34$ & $1.70$ & $2.17$ & $3.60$ & $4.74$ & $22.96$ \\
HD 50138 & $2.13$ & $4.29$ & $10.45$ & $26.11$ & - & $444.08$ & $2375.59$ \\
HD 56895B & $1.12$ & $1.07$ & $1.19$ & - & - & - & - \\
HD 58647 & $1.29$ & $1.73$ & $3.27$ & $8.58$ & $27.12$ & $39.00$ & $105.20$ \\
HD 59319 & $1.15$ & $1.11$ & $1.16$ & $1.24$ & $1.21$ & $0.88$ & $129.98$ \\
HD 68695 & $1.53$ & $2.61$ & $5.20$ & $15.95$ & $26.36$ & $96.34$ & $1314.61$ \\
HD 76534 & $1.31$ & $1.31$ & $1.40$ & $2.21$ & $2.55$ & $3.90$ & $29.05$ \\
HD 85567 & $1.87$ & $3.74$ & $8.55$ & $22.94$ & $58.45$ & $209.67$ & $836.94$ \\
HD 87403 & $1.35$ & $1.36$ & $1.49$ & $1.67$ & $1.76$ & $1.15$ & $10.21$ \\
HD 87643 & $3.76$ & $13.14$ & $41.40$ & - & - & $2278.72$ & $12716.58$ \\
HD 94509 & $1.28$ & $1.40$ & $1.53$ & $2.17$ & $2.55$ & $3.48$ & $17.17$ \\
HD 95881 & $2.28$ & $4.48$ & $10.93$ & $33.15$ & $86.18$ & $295.79$ & $1007.97$ \\
HD 96042 & $1.47$ & $1.45$ & $1.71$ & $1.48$ & $1.49$ & $1.66$ & $45.46$ \\
HD 9672 & $1.04$ & $1.00$ & $1.12$ & $1.05$ & $1.22$ & $0.79$ & $3.97$ \\
HD 97048 & $1.84$ & $3.01$ & $5.88$ & $19.83$ & $20.04$ & $239.06$ & $3219.89$ \\
HD 98922 & $2.15$ & $4.46$ & $11.09$ & - & - & $352.13$ & $1148.15$ \\
HR 5999 & $1.59$ & $2.82$ & $6.25$ & - & - & $112.50$ & $414.85$ \\
HT CMa & $4.29$ & $9.27$ & $20.90$ & $64.36$ & $148.82$ & $983.40$ & $5396.58$ \\
HU CMa & $1.58$ & $3.04$ & $7.04$ & $20.04$ & $42.48$ & $535.48$ & $3456.52$ \\
Hen 3-1121 & $1.73$ & $4.13$ & $10.01$ & $26.17$ & $62.07$ & $699.79$ & $6435.45$ \\
Hen 3-1121S & $0.87$ & $0.82$ & $0.85$ & $0.87$ & $1.24$ & $29.55$ & $1962.39$ \\
Hen 3-1191 & $4.48$ & $21.19$ & $98.70$ & $510.27$ & $1995.88$ & $10315.68$ & $37787.17$ \\
Hen 3-823 & $1.35$ & $1.45$ & $1.73$ & $2.15$ & $2.63$ & $3.62$ & $9.41$ \\
Hen 3-847 & $1.38$ & $2.77$ & $12.52$ & $127.89$ & $761.32$ & $9538.41$ & $58854.42$ \\
Hen 3-938 & $2.93$ & $8.65$ & $27.33$ & $69.17$ & $213.51$ & $823.51$ & $3658.69$ \\
IL Cep & $0.85$ & $0.84$ & $0.89$ & $0.92$ & $1.12$ & $1.12$ & - \\
IP Per & $1.80$ & $3.18$ & $6.85$ & $15.40$ & $27.39$ & $36.94$ & $645.90$ \\
KK Oph & $3.04$ & $12.33$ & $41.94$ & $102.96$ & $243.95$ & $1149.69$ & $4781.29$ \\
LKHa 260 & $1.49$ & $3.11$ & - & $22.31$ & $53.84$ & $388.18$ & $312.56$ \\
LKHa 338 & $4.59$ & $11.05$ & $38.14$ & $193.55$ & $635.30$ & $6577.01$ & $38467.51$ \\
LkHa 208 & $1.73$ & $2.31$ & $4.08$ & $12.66$ & $43.63$ & $1204.97$ & $7023.91$ \\
LkHa 215 & $2.25$ & $4.00$ & $7.74$ & $27.52$ & $53.10$ & $183.08$ & $1801.94$ \\
LkHa 257 & $1.55$ & $2.63$ & $4.83$ & $8.08$ & $13.28$ & $58.14$ & $1381.77$ \\
LkHa 259 & $2.90$ & $3.65$ & $5.33$ & $10.63$ & $24.83$ & $580.94$ & $10317.88$ \\
LkHa 324 & - & $1.06$ & $1.13$ & $1.61$ & $2.73$ & $73.32$ & $653.45$ \\
LkHa 339 & $0.75$ & $1.07$ & $2.15$ & $5.96$ & $13.88$ & $243.18$ & $2014.27$ \\
MQ Cas & $5.39$ & $19.10$ & $58.45$ & - & - & - & - \\
MWC 1021 & $1.54$ & $1.76$ & $2.77$ & - & - & $37.54$ & $74.27$ \\
MWC 1080 & $3.17$ & $8.57$ & $20.05$ & - & - & $579.77$ & $2359.22$ \\
MWC 137 & $1.94$ & $3.25$ & $8.25$ & $30.58$ & $86.52$ & $328.65$ & $3052.44$ \\
MWC 297 & $1.55$ & $3.71$ & $8.52$ & - & - & $280.89$ & $4022.76$ \\
MWC 342 & $2.99$ & $6.55$ & $15.48$ & - & - & $717.92$ & $3665.94$ \\
MWC 593 & $1.61$ & $1.83$ & $2.24$ & - & - & - & - \\
MWC 655 & $3.26$ & $4.01$ & $6.05$ & $5.07$ & $6.92$ & $41.93$ & $1084.04$ \\
MWC 657 & $1.41$ & $2.98$ & $6.76$ & $18.91$ & $66.61$ & $242.44$ & $800.51$ \\
MWC 878 & $1.61$ & $2.72$ & $7.30$ & $43.34$ & $157.49$ & $976.37$ & $2629.98$ \\
MWC 953 & $1.26$ & $1.42$ & $1.65$ & $1.99$ & $2.20$ & $3.13$ & $102.09$ \\
NSV 2968 & $2.31$ & $4.35$ & $10.22$ & $12.67$ & $41.90$ & $297.56$ & $2246.08$ \\
NV Ori & $1.19$ & $1.64$ & $2.92$ & $5.05$ & $8.88$ & $88.23$ & $553.15$ \\
PDS 002 & $1.08$ & $1.24$ & $1.76$ & $3.39$ & $4.95$ & $51.00$ & $718.19$ \\
PDS 004 & $1.43$ & $2.17$ & $3.79$ & $7.57$ & $15.55$ & $160.54$ & $1084.87$ \\
PDS 021 & $1.93$ & $4.07$ & $8.97$ & $18.07$ & $41.18$ & $595.37$ & $4130.11$ \\
PDS 022 & $1.21$ & $1.48$ & $2.47$ & $6.19$ & $13.77$ & $340.58$ & $4104.11$ \\
PDS 025 & $3.07$ & $6.73$ & $15.36$ & $37.43$ & $63.06$ & $373.77$ & $3752.05$ \\
PDS 123 & $2.58$ & $6.31$ & $14.05$ & $32.32$ & $65.17$ & $326.20$ & $2513.33$ \\
PDS 124 & $1.69$ & $2.51$ & $4.95$ & $15.08$ & $30.21$ & $456.06$ & $2857.71$ \\
PDS 126 & $1.51$ & $2.16$ & $4.09$ & $8.08$ & $14.23$ & $79.11$ & $390.36$ \\
PDS 129 & $1.38$ & $1.58$ & $2.17$ & $3.86$ & $6.50$ & $78.22$ & $630.91$ \\
PDS 130 & $2.13$ & $4.13$ & $8.84$ & $22.69$ & $45.11$ & $500.19$ & $3967.51$ \\
PDS 133 & $2.13$ & $6.27$ & $16.14$ & $23.84$ & $77.26$ & $645.47$ & $6061.49$ \\
PDS 134 & $1.79$ & $1.90$ & $2.09$ & - & - & - & - \\
PDS 138 & $1.20$ & $1.20$ & $1.24$ & - & - & - & - \\
PDS 174 & $2.24$ & $3.43$ & $5.38$ & $12.27$ & $26.19$ & $537.78$ & $29319.27$ \\
PDS 211 & $1.35$ & $2.16$ & $4.29$ & $7.32$ & $18.45$ & $433.58$ & $3154.52$ \\
PDS 24 & $2.15$ & $3.93$ & $8.37$ & $18.41$ & $38.89$ & $852.57$ & $9838.27$ \\
PDS 241 & $1.30$ & $1.51$ & $1.86$ & $7.04$ & $19.17$ & $5641.92$ & $83493.60$ \\
PDS 27 & $1.78$ & $4.54$ & $12.10$ & $31.59$ & $134.49$ & $1091.96$ & $7522.03$ \\
PDS 277 & $1.23$ & $1.54$ & $2.80$ & $6.42$ & $10.08$ & $70.34$ & $1354.70$ \\
PDS 286 & $1.73$ & $1.85$ & $2.39$ & $0.99$ & $1.33$ & $1.82$ & $21.31$ \\
PDS 290 & $0.64$ & $0.67$ & $0.76$ & $0.63$ & $0.61$ & $1.88$ & $111.26$ \\
PDS 297 & $1.12$ & $1.12$ & $1.15$ & $1.13$ & $1.16$ & $1.11$ & $18.69$ \\
PDS 324 & $1.34$ & $1.53$ & $1.79$ & - & - & - & - \\
PDS 33 & $1.80$ & $2.83$ & $4.94$ & $11.66$ & $21.41$ & $458.11$ & $4167.59$ \\
PDS 34 & $2.61$ & $5.93$ & $14.05$ & $31.75$ & $63.65$ & $605.75$ & $7602.75$ \\
PDS 344 & $1.78$ & $2.17$ & $2.65$ & $4.71$ & $8.69$ & $390.97$ & $3434.61$ \\
PDS 361S & $2.87$ & $3.39$ & $3.74$ & $5.80$ & $6.63$ & $4.67$ & - \\
PDS 37 & $0.71$ & $2.38$ & $6.48$ & $18.45$ & $55.76$ & $526.99$ & $4328.48$ \\
PDS 389 & $1.55$ & $2.02$ & $3.23$ & $4.82$ & $8.86$ & $38.90$ & $328.32$ \\
PDS 415N & $0.92$ & $1.14$ & $1.62$ & $4.08$ & $9.07$ & $21.21$ & $332.90$ \\
PDS 431 & $1.41$ & $1.41$ & $1.55$ & $1.50$ & $1.54$ & $31.35$ & - \\
PDS 469 & $1.44$ & $1.85$ & $2.61$ & $4.02$ & $6.10$ & $214.67$ & $1185.71$ \\
PDS 477 & $3.47$ & $9.52$ & $24.09$ & $61.85$ & $129.93$ & $1201.43$ & $9937.68$ \\
PDS 520 & $4.96$ & $8.63$ & $18.99$ & $41.85$ & $94.93$ & $429.65$ & $3731.82$ \\
PDS 543 & $0.71$ & $0.62$ & $0.65$ & $0.76$ & $1.94$ & $19.23$ & $572.04$ \\
PDS 69 & $1.15$ & $2.19$ & $4.42$ & - & - & - & - \\
PX Vul & $1.75$ & $2.68$ & $4.57$ & $10.61$ & $18.18$ & $46.64$ & $341.19$ \\
RR Tau & $1.23$ & $3.42$ & $8.44$ & $24.63$ & $48.23$ & $156.34$ & $996.01$ \\
RY Ori & $1.35$ & $1.63$ & $2.70$ & $4.49$ & $7.27$ & $58.82$ & $358.89$ \\
SAO 185668 & $0.87$ & $0.85$ & $0.84$ & $0.82$ & $1.01$ & $25.56$ & $881.06$ \\
SAO 220669 & $1.43$ & $1.43$ & $1.46$ & $1.65$ & $1.56$ & $3.78$ & $251.64$ \\
SV Cep & $2.62$ & $5.05$ & $10.71$ & $23.65$ & $49.17$ & $743.44$ & $4567.70$ \\
T Ori & $2.65$ & $5.88$ & $14.38$ & $35.54$ & $52.41$ & $192.59$ & $1528.40$ \\
TY CrA & $1.59$ & $2.16$ & $2.66$ & $5.32$ & $14.92$ & $3250.79$ & $6107.38$ \\
UX Ori & $2.89$ & $4.96$ & $10.85$ & $23.75$ & $46.70$ & $338.18$ & $2472.27$ \\
V1012 Ori & $1.42$ & $2.75$ & $5.85$ & $10.93$ & $20.05$ & $94.49$ & $1839.69$ \\
V1295 Aql & $1.30$ & $2.07$ & $4.38$ & $11.48$ & $29.27$ & $118.54$ & $396.22$ \\
V1478 Cyg & $2.70$ & $4.71$ & $13.20$ & - & - & $202.01$ & $860.06$ \\
V1493 Cyg & $2.39$ & $3.92$ & $7.04$ & $16.28$ & $35.08$ & $197.28$ & $900.73$ \\
V1685 Cyg & $1.67$ & $3.51$ & $7.74$ & - & - & - & $3124.43$ \\
V1686 Cyg & $3.11$ & $9.36$ & $28.85$ & - & - & - & - \\
V1787 Ori & $2.03$ & $3.14$ & $6.49$ & $15.63$ & $33.22$ & $236.07$ & $1369.96$ \\
V1818 Ori & $0.36$ & $1.94$ & $7.77$ & $28.86$ & $86.74$ & $453.29$ & $2698.23$ \\
V1977 Cyg & $2.09$ & $4.14$ & $10.33$ & $25.91$ & $66.33$ & $153.92$ & $794.03$ \\
V2019 Cyg & $1.36$ & $1.88$ & $3.05$ & $6.09$ & $10.78$ & - & $6223.77$ \\
V346 Ori & $1.14$ & $1.69$ & $3.15$ & $9.45$ & $15.76$ & $44.81$ & $776.76$ \\
V350 Ori & $1.28$ & $2.45$ & $5.47$ & $11.91$ & $20.34$ & $133.53$ & $1164.99$ \\
V351 Ori & $1.73$ & $2.44$ & $4.68$ & $11.46$ & $22.68$ & $44.22$ & $691.22$ \\
V361 Cep & $1.08$ & $1.20$ & $1.42$ & $2.00$ & $2.93$ & $6.86$ & $1741.54$ \\
V373 Cep & $3.30$ & $8.69$ & $21.53$ & $139.91$ & $732.16$ & $2133.92$ & $42216.18$ \\
V374 Cep & $1.57$ & $1.83$ & $2.42$ & $1.98$ & $2.76$ & $4.93$ & $27.19$ \\
V380 Ori & $2.31$ & $5.36$ & $12.67$ & $27.79$ & $74.38$ & $387.31$ & $1548.31$ \\
V388 Vel & $1.34$ & $3.55$ & $9.33$ & $40.19$ & $94.40$ & $1665.31$ & $21691.47$ \\
V431 Sct & $1.34$ & $2.96$ & $15.27$ & $129.70$ & - & $4848.99$ & $34190.80$ \\
V594 Cas & $3.39$ & $8.94$ & $21.62$ & $49.46$ & $113.26$ & $522.05$ & $2764.98$ \\
V594 Cyg & $26.23$ & $29.73$ & $43.29$ & $25.55$ & $34.83$ & $54.10$ & $169.97$ \\
V599 Ori & $1.70$ & $2.62$ & $5.27$ & $6.94$ & $12.85$ & $41.44$ & $959.30$ \\
V669 Cep & $1.83$ & $3.48$ & $10.37$ & $44.40$ & $174.02$ & $2080.63$ & $9413.59$ \\
V718 Sco & $1.72$ & $2.24$ & $3.81$ & $7.66$ & $16.83$ & $155.15$ & $707.61$ \\
V921 Sco & $3.58$ & $8.44$ & $24.92$ & - & - & $3378.91$ & $57901.14$ \\
VV Ser & $3.57$ & $8.83$ & $21.97$ & $46.08$ & $121.07$ & $428.03$ & $1766.30$ \\
VX Cas & $2.00$ & $4.62$ & $11.25$ & $27.13$ & $45.17$ & $276.95$ & $2735.55$ \\
WRAY 15-1435 & $2.24$ & $2.88$ & $3.34$ & $4.39$ & $7.36$ & $192.05$ & $3602.51$ \\
WW Vul & $2.19$ & $4.60$ & $10.53$ & $17.32$ & $32.02$ & $231.68$ & $1366.23$ \\
XY Per A & $2.61$ & $4.51$ & $9.33$ & $21.53$ & $47.11$ & $131.33$ & $670.35$ \\
\hline
\end{longtable}

\newpage

\begin{landscape}
\renewcommand{\arraystretch}{1.2}
\begin{longtable}{lrrrrrrrrc}
\caption{\label{table2_2} Main parameters of each Herbig Ae/Be star belonging to the low quality sample of 34 sources.}\\
\hline\hline 
Name & RA &  DEC & Parallax & Distance &T\textsubscript{eff} &Log(L) & A\textsubscript{V} & V & Binary\\
  & (h:m:s)  & (deg:m:s) &  (mas) & (pc) & (K)  & ($L_{\odot}$) & (mag) & (mag) & \\ 
\hline
\endfirsthead
\caption{continued.}\\
\hline\hline
Name & RA &  DEC & Parallax & Distance &T\textsubscript{eff} &Log(L) & A\textsubscript{V} & V & Binary\\
  & (h:m:s)  & (deg:m:s) &  (mas) & (pc) & (K)  & ($L_{\odot}$) & (mag) & (mag) & \\ 
\hline
\endhead
\hline
\endfoot
BP Psc & 23:22:24.7 & -02:13:42 & $2.79\pm{0.39}$ & $350^{+110}_{-50}$ & $5350^{+80}_{-70}$ & $0.73^{+0.34}_{-0.23}$ & $0.83^{+0.25}_{-0.24}$ & 11.53 & - \\
DK Cha & 12:53:17.1 & -77:07:11 & $4.10\pm{0.37}$ & $243^{+47}_{-28}$ & $7250^{+130}_{-130}$ & $0.47^{+0.20}_{-0.16}$ & $8.12^{+0.11}_{-0.14}$ & 18.54 & - \\
GSC 5360-1033 & 05:57:49.5 & -14:05:34 & $1.649\pm{0.034}$ & $605^{+22}_{-19}$ & $15000^{+800}_{-1000}$ & $1.01^{+0.27}_{-0.30}$ & $1.60^{+0.50}_{-0.50}$ & 13.91 & Yes$^{29}$ \\
GSC 5988-2257 & 07:41:41.1 & -20:00:13 & $-4.66\pm{0.86}$ & $980^{+520}_{-270}$ & $16500^{+3000}_{-800}$ & $1.52^{+0.63}_{-0.39}$ & $3.18^{+0.23}_{-0.15}$ & 15.52 & - \\
GSC 6542-2339 & 07:24:37.0 & -24:34:47 & $1.12\pm{0.11}$ & $850^{+160}_{-100}$ & $32900^{+2000}_{-3900}$ & $3.03^{+0.27}_{-0.29}$ & $5.24^{+0.14}_{-0.18}$ & 15.12 & - \\
HBC 1 & 00:07:02.6 & +65:38:38 & $0.16\pm{0.52}$ & $760^{+440}_{-190}$ & $8150^{+180}_{-160}$ & $-0.65^{+0.47}_{-0.32}$ & $1.05^{+0.19}_{-0.16}$ & 16.76 & - \\
HBC 324 & 00:07:30.7 & +65:39:53 & $-0.27\pm{0.47}$ & $900^{+470}_{-220}$ & $7830^{+160}_{-220}$ & $0.80^{+0.47}_{-0.35}$ & $2.51^{+0.25}_{-0.25}$ & 14.94 & - \\
HBC 694 & 20:24:29.5 & +42:14:02 & $0.90\pm{0.43}$ & $670^{+370}_{-150}$ & $8150^{+180}_{-160}$ & $-0.50^{+0.58}_{-0.43}$ & $1.93^{+0.50}_{-0.50}$ & 17.00 & - \\
HBC 717 & 20:52:06.0 & +44:17:16 & $0.49\pm{0.12}$ & $1390^{+390}_{-220}$ & $6400^{+150}_{-150}$ & $1.88^{+0.32}_{-0.23}$ & $2.82^{+0.27}_{-0.20}$ & 13.55 & - \\
HD 245906 & 05:39:30.5 & +26:19:55 & $0.67\pm{0.48}$ & $690^{+400}_{-170}$ & $7990^{+160}_{-160}$ & $1.66^{+0.46}_{-0.34}$ & $0.85^{+0.16}_{-0.25}$ & 10.55 & Yes$^{2}$ \\
HD 53367 & 07:04:25.5 & -10:27:16 & $7.77\pm{0.79}$ & $130^{+30}_{-17}$ & $29500^{+1000}_{-1000}$ & $3.13^{+0.23}_{-0.17}$ & $2.051^{+0.043}_{-0.050}$ & 7.36 & Yes$^{4}$ \\
HD 72106B & 08:29:34.9 & -38:36:21 & $0.03\pm{0.83}$ & $600^{+430}_{-170}$ & $8750^{+250}_{-250}$ & $1.85^{+0.53}_{-0.38}$ & $0.51^{+0.13}_{-0.21}$ & 9.50 & Yes$^{20}$ \\
Hen 2-80 & 12:22:23.2 & -63:17:17 & $0.71\pm{0.38}$ & $750^{+390}_{-170}$ & $14000^{+1000}_{-1000}$ & $2.12^{+0.49}_{-0.37}$ & $2.97^{+0.16}_{-0.18}$ & 12.79 & - \\
MWC 314 & 19:21:34.0 & +14:52:57 & $0.191\pm{0.042}$ & $2980^{+550}_{-370}$ & $16500^{+3000}_{-800}$ & $5.29^{+0.52}_{-0.37}$ & $4.50^{+0.50}_{-0.50}$ & 9.80 & Yes$^{31}$ \\
MWC 623 & 19:56:31.5 & +31:06:20 & $0.173\pm{0.036}$ & $3280^{+570}_{-390}$ & $15800^{+1000}_{-1000}$ & $4.58^{+0.28}_{-0.25}$ & $3.77^{+0.19}_{-0.17}$ & 10.92 & Yes$^{21}$ \\
MWC 930 & 18:26:25.2 & -07:13:18 & $-0.162\pm{0.094}$ & $2590^{+650}_{-420}$ & $11900^{+1700}_{-1400}$ & $5.85^{+0.39}_{-0.39}$ & $8.72^{+0.18}_{-0.31}$ & 11.51 & - \\
NX Pup & 07:19:28.3 & -44:35:11 & $-9.84\pm{0.65}$ & $1670^{+630}_{-380}$ & $7000^{+250}_{-250}$ & $2.46^{+0.30}_{-0.22}$ & $0.000^{+0.070}_{-0.000}$ & 9.63 & Yes$^{4}$ \\
PDS 144S & 15:49:15.3 & -26:00:55 & $6.69\pm{0.12}$ & $149.6^{+4.6}_{-4.2}$ & $7750^{+250}_{-250}$ & $-0.673^{+0.057}_{-0.057}$ & $0.570^{+0.070}_{-0.080}$ & 12.79 & - \\
PDS 229N & 06:55:40.0 & -03:09:50 & $-0.52\pm{0.53}$ & $880^{+470}_{-220}$ & $12500^{+250}_{-250}$ & $1.70^{+0.44}_{-0.32}$ & $2.13^{+0.12}_{-0.12}$ & 13.07 & Yes$^{22}$ \\
PDS 322 & 10:52:08.7 & -56:12:07 & $-1.30\pm{0.27}$ & $1730^{+610}_{-360}$ & $19500^{+5000}_{-3000}$ & $2.88^{+0.60}_{-0.47}$ & $1.39^{+0.29}_{-0.23}$ & 11.97 & - \\
PDS 364 & 13:20:03.6 & -62:23:54 & $-0.39\pm{0.12}$ & $2430^{+660}_{-420}$ & $12500^{+1000}_{-1000}$ & $2.32^{+0.31}_{-0.26}$ & $1.870^{+0.050}_{-0.030}$ & 13.46 & - \\
PDS 371 & 13:47:31.4 & -36:39:50 & $9.87\pm{0.21}$ & $101.4^{+3.8}_{-3.4}$ & $32900^{+2000}_{-3900}$ & $0.98^{+0.15}_{-0.21}$ & $5.16^{+0.14}_{-0.18}$ & 15.55 & - \\
PDS 453 & 17:20:56.1 & -26:03:31 & $5.70\pm{0.57}$ & $176^{+40}_{-22}$ & $7000^{+120}_{-250}$ & $-0.43^{+0.27}_{-0.23}$ & $1.44^{+0.25}_{-0.29}$ & 13.41 & - \\
PDS 530 & 18:41:34.4 & +08:08:21 & $0.27\pm{0.17}$ & $1390^{+470}_{-260}$ & $8150^{+180}_{-160}$ & $1.35^{+0.33}_{-0.24}$ & $1.53^{+0.19}_{-0.16}$ & 13.54 & - \\
PDS 551 & 18:55:23.0 & +04:04:35 & $1.99\pm{0.12}$ & $496^{+58}_{-42}$ & $29000^{+3900}_{-4500}$ & $0.93^{+0.41}_{-0.43}$ & $2.90^{+0.50}_{-0.50}$ & 16.60 & - \\
PDS 581 & 19:36:18.9 & +29:32:50 & $0.96\pm{0.38}$ & $690^{+350}_{-150}$ & $24500^{+4500}_{-5000}$ & $2.89^{+0.62}_{-0.56}$ & $2.63^{+0.26}_{-0.29}$ & 11.75 & - \\
PV Cep & 20:45:54.0 & +67:57:39 & $2.910\pm{0.059}$ & $343^{+12}_{-11}$ & $8150^{+180}_{-160}$ & $0.00^{+0.11}_{-0.09}$ & $5.12^{+0.19}_{-0.16}$ & 17.46 & - \\
R CrA & 19:01:53.7 & -36:57:09 & $10.54\pm{0.70}$ & $95^{+13}_{-9}$ & $8150^{+180}_{-160}$ & $-0.06^{+0.19}_{-0.15}$ & $2.13^{+0.19}_{-0.16}$ & 11.85 & - \\
UY Ori & 05:32:00.3 & -04:55:54 & $2.811\pm{0.082}$ & $355^{+18}_{-16}$ & $9750^{+250}_{-250}$ & $0.394^{+0.072}_{-0.059}$ & $1.110^{+0.020}_{-0.000}$ & 12.79 & - \\
V590 Mon & 06:40:44.6 & +09:48:02 & $1.14\pm{0.13}$ & $820^{+170}_{-100}$ & $12500^{+1000}_{-1000}$ & $1.38^{+0.26}_{-0.21}$ & $1.030^{+0.040}_{-0.050}$ & 12.60 & Yes$^{2}$ \\
V645 Cyg & 21:39:58.3 & +50:14:21 & $0.53\pm{0.40}$ & $790^{+410}_{-180}$ & $36900^{+2000}_{-2000}$ & $3.60^{+0.48}_{-0.34}$ & $4.51^{+0.12}_{-0.12}$ & 13.10 & - \\
V892 Tau & 04:18:40.6 & +28:19:15 & $8.52\pm{0.12}$ & $117.5^{+2.7}_{-2.5}$ & $11500^{+1500}_{-800}$ & $0.13^{+0.33}_{-0.28}$ & $4.87^{+0.50}_{-0.50}$ & 15.17 & Yes$^{5}$ \\
VY Mon & 06:31:06.9 & +10:26:05 & $-1.94\pm{0.38}$ & $1470^{+580}_{-330}$ & $12000^{+4000}_{-4000}$ & $3.56^{+0.64}_{-0.68}$ & $5.68^{+0.17}_{-0.45}$ & 12.97 & - \\
Z CMa & 07:03:43.2 & -11:33:06 & $4.30\pm{0.89}$ & $230^{+150}_{-50}$ & $8500^{+500}_{-500}$ & $2.25^{+0.51}_{-0.29}$ & $3.37^{+0.12}_{-0.16}$ & 9.25 & Yes$^{1}$ \\
\hline
\end{longtable}
\tablebib{Atmospheric parameters T\textsubscript{eff}, A\textsubscript{V} and V taken from the following sources in order of choice: \citealp{Fairlamb}; \citealp{Montesinos}; \citealp{Hernandez}; \citealp{Mendigutia}; \citealp{Carmona}; \citealp{Chen}; \citealp{Alecian}; \citealp{Sartori}; \citealp{Manoj}; \citealp{Hernandez2}; \citealp{Vieira}; APASS Data Release 9 and the SIMBAD database. If not available they were derived as described in Sect. \ref{sec:Atmospheric}. See Sect. \ref{sec:Der_cuantities} for derivation of $L$, Mass and Age. The references for binarity are: (1) \citet{Baines}; (2) \citet{Wheelwright}; (3) \citet{Leinert}; (4) \citet{Maheswar}; (5) \citet{Wheelwright2}; (6) \citet{Alecian}; (7) \citet{Hamaguchi}; (8) \citet{Dunhill}; (9) \citet{Coulson and Walther}; (10) \citet{Liu}; (11) \citet{Biller}; (12) \citet{Schuetz}; (13) \citet{Boersma}; (14) \citet{Malkov}; (15) \citet{Ferro}; (16) \citet{Kubat}; (17) \citet{Morrell}; (18) \citet{Lazareff}; (19) \citet{Mayer}; (20) \citet{Folsom}; (21) \citet{Corporon}; (22) \citet{Doering}; (23) \citet{Chelli}; (24) \citet{Miroshnichenko}; (25) \citet{Friedemann}; (26) \citet{Kraus}; (27) \citet{Torres}; (28) \citet{Aspin}; (29) \citet{Connelley}; (30) \citet{Millour}; (31) \citet{Frasca}; (32) \citet{Marston}; (33) \citet{Zhang3}.}
\end{landscape}

\newpage

\begin{landscape}
\renewcommand{\arraystretch}{1.2}
\begin{longtable}{lrrrrrrr}
\caption{\label{table3_2} Other parameters of each Herbig Ae/Be star belonging to the low quality sample of 34 sources.}\\
\hline\hline 
Name & Near IR excess &  Mid IR excess & H$\alpha$ EW & H$\alpha$ & $V_{i}$ & Mass & Age\\
  & ($1.24-3.4\mu$m)  & ($3.4-22\mu$m) &  (\AA) & line shape & & ($M_{\odot}$) & (Myr)\\ 
\hline
\endfirsthead
\caption{continued.}\\
\hline\hline
Name & Near IR excess &  Mid IR excess & H$\alpha$ EW & H$\alpha$ & $V_{i}$ & Mass & Age\\
  & ($1.24-3.4\mu$m)  & ($3.4-22\mu$m) &  (\AA) & line shape & & ($M_{\odot}$) & (Myr)\\ 
\hline
\endhead
\hline
\endfoot
BP Psc & $0.46^{+0.24}_{-0.19}$ & $0.81^{+0.27}_{-0.21}$ & $-14.83\pm{0.35}^{33}$ & - & - & $1.90^{+0.50}_{-0.26}$ & $1.7^{+1.6}_{-1.0}$ \\
DK Cha & $8.0^{+2.2}_{-1.5}$ & $12.6^{+4.5}_{-3.0}$ & $-95.3\pm{4.4}^{24}$ & - & - & $1.369^{+0.068}_{-0.068}$ & $17.2^{+2.8}_{-3.6}$ \\
GSC 5360-1033 & - & - & $-9.36\pm{0.20}^{15}$ & d$^{21}$ & - & - & - \\
GSC 5988-2257 & $0.077^{+0.028}_{-0.037}$ & $0.147^{+0.049}_{-0.068}$ & $-19.83\pm{0.75}^{27}$ & d$^{21}$ & - & - & - \\
GSC 6542-2339 & - & - & $-28.0\pm{1.2}^{15}$ & d$^{21}$ & - & - & - \\
HBC 1 & $83^{+19}_{-17}$ & $237^{+58}_{-51}$ & $-40.2\pm{3.1}^{29}$ & - & - & - & - \\
HBC 324 & $0.124^{+0.062}_{-0.053}$ & $1.11^{+0.31}_{-0.24}$ & $-26.16\pm{0.84}^{5}$ & - & - & $1.50^{+0.29}_{-0.07}$ & $13.2^{+6.8}_{-7.3}$ \\
HBC 694 & $0.99^{+0.68}_{-0.41}$ & $2.9^{+1.9}_{-1.1}$ & - & - & - & - & - \\
HBC 717 & $0.32^{+0.12}_{-0.12}$ & $0.44^{+0.11}_{-0.11}$ & $-23.59\pm{0.95}^{5}$ & - & - & $3.4^{+1.0}_{-0.5}$ & $0.88^{+0.52}_{-0.50}$ \\
HD 245906 & $0.145^{+0.066}_{-0.037}$ & $0.091^{+0.027}_{-0.015}$ & $-12.80\pm{0.18}^{5}$ & P$^{26}$ & - & $2.4^{+1.1}_{-0.5}$ & $2.8^{+2.7}_{-1.8}$ \\
HD 53367 & $(0.79^{+0.27}_{-0.21})\cdot10^{-3}$ & $(0.42^{+0.11}_{-0.08})\cdot10^{-3}$ & $-14.00\pm{0.70}^{1}$ & d$^{21}$ & - & - & - \\
HD 72106B & $0.077^{+0.033}_{-0.020}$ & $0.114^{+0.031}_{-0.018}$ & $-5.8\pm{1.6}^{1}$ & s$^{21}$ & - & $2.7^{+1.5}_{-0.7}$ & $2.1^{+2.6}_{-1.5}$ \\
Hen 2-80 & $0.146^{+0.070}_{-0.045}$ & $0.51^{+0.24}_{-0.15}$ & $-155.5\pm{7.5}^{2}$ & d$^{2}$ & - & $3.07^{+0.90}_{-0.39}$ & $2.2^{+8.7}_{-1.3}$ \\
MWC 314 & $0.017^{+0.021}_{-0.012}$ & $(0.27^{+0.32}_{-0.18})\cdot10^{-2}$ & $-130\pm{15}^{35}$ & d$^{36}$ & - & $24.6^{+1.2}_{-1.2}$ & $(10.00^{+0.80}_{-0.50})\cdot10^{-3}$ \\
MWC 623 & $0.084^{+0.042}_{-0.028}$ & $0.057^{+0.043}_{-0.024}$ & $-134\pm{14}^{16}$ & s$^{16}$ & - & $18.2^{+4.7}_{-3.1}$ & $0.015^{+0.010}_{-0.005}$ \\
MWC 930 & - & - & - & - & - & - & - \\
NX Pup & $0.78^{+0.11}_{-0.15}$ & $1.25^{+0.26}_{-0.27}$ & $-54.0\pm{3.0}^{1}$ & d$^{8}$ & -* & $5.0^{+1.4}_{-0.7}$ & $0.28^{+0.20}_{-0.17}$ \\
PDS 144S & $2.88^{+0.34}_{-0.29}$ & $6.0^{+1.1}_{-0.9}$ & $-29.2\pm{1.5}^{1}$ & s$^{31}$ & - & - & - \\
PDS 229N & $0.044^{+0.012}_{-0.010}$ & $0.057^{+0.011}_{-0.009}$ & $-2.2\pm{1.3}^{1}$ & P$^{21}$ & - & $2.51^{+0.47}_{-0.13}$ & $5^{+13}_{-3}$ \\
PDS 322 & $(0.23^{+0.34}_{-0.23})\cdot10^{-2}$ & $0.029^{+0.025}_{-0.016}$ & $3.792\pm{0.020}^{15}$ & - & - & $5.4^{+2.6}_{-1.5}$ & $1.1^{+5.3}_{-0.9}$ \\
PDS 364 & $0.104^{+0.026}_{-0.023}$ & $0.227^{+0.057}_{-0.049}$ & $-88.0\pm{1.2}^{1}$ & d$^{21}$ & - & $3.30^{+0.90}_{-0.50}$ & $1.45^{+0.75}_{-0.75}$ \\
PDS 371 & $(0.53^{+0.33}_{-0.16})\cdot10^{-2}$ & $(0.66^{+0.36}_{-0.17})\cdot10^{-2}$ & $-43.0\pm{2.0}^{27}$ & s$^{21}$ & - & - & - \\
PDS 453 & $0.50^{+0.22}_{-0.15}$ & $0.73^{+0.25}_{-0.17}$ & $-1.42\pm{0.20}^{15}$ & d$^{21}$ & -* & - & - \\
PDS 530 & $0.47^{+0.10}_{-0.10}$ & $1.37^{+0.25}_{-0.25}$ & $-37.2\pm{1.4}^{15}$ & s$^{21}$ & - & $1.89^{+0.52}_{-0.23}$ & $5.2^{+2.2}_{-2.4}$ \\
PDS 551 & $0.34^{+0.45}_{-0.18}$ & $0.8^{+1.0}_{-0.4}$ & $-51.2\pm{2.5}^{15}$ & d$^{21}$ & - & - & - \\
PDS 581 & $0.061^{+0.087}_{-0.030}$ & $0.25^{+0.43}_{-0.14}$ & $-201\pm{10}^{15}$ & s$^{21}$ & - & $5.4^{+2.9}_{-0.3}$ & $0.6^{+1.1}_{-0.4}$ \\
PV Cep & $4.7^{+1.2}_{-1.0}$ & $24.7^{+6.6}_{-5.7}$ & $-59.1\pm{2.5}^{28}$ & P$^{28}$ & - & - & - \\
R CrA & $16.6^{+7.1}_{-5.2}$ & $23^{+12}_{-8}$ & $-94.7\pm{4.3}^{17}$ & d$^{26}$ & - & - & - \\
UY Ori & $0.231^{+0.018}_{-0.021}$ & $0.930^{+0.058}_{-0.072}$ & $-10.3\pm{1.6}^{1}$ & P$^{31}$ & - & - & - \\
V590 Mon & $0.131^{+0.037}_{-0.028}$ & $0.53^{+0.15}_{-0.11}$ & $-69.7\pm{1.1}^{1}$ & d$^{31}$ & - & $2.30^{+0.13}_{-0.11}$ & $6^{+14}_{-1}$ \\
V645 Cyg & $(0.97^{+0.39}_{-0.28})\cdot10^{-2}$ & $0.119^{+0.078}_{-0.046}$ & $-125\pm{12}^{17}$ & d$^{28}$ & - & - & - \\
V892 Tau & $1.7^{+1.6}_{-0.9}$ & $3.9^{+3.6}_{-2.1}$ & $-24.47\pm{0.89}^{5}$ & - & - & - & - \\
VY Mon & $0.17^{+0.37}_{-0.10}$ & $0.32^{+0.79}_{-0.20}$ & $-26.3\pm{2.0}^{1}$ & P$^{31}$ & -* & $8.8^{+7.5}_{-4.0}$ & $0.08^{+0.49}_{-0.07}$ \\
Z CMa & $0.25^{+0.11}_{-0.07}$ & $0.43^{+0.23}_{-0.15}$ & $-25.0\pm{7.0}^{1}$ & P$^{8}$ & - & $3.8^{+2.0}_{-0.8}$ & $0.80^{+0.83}_{-0.59}$ \\
\hline
\end{longtable}
\tablebib{The H$\alpha$ line profile classification is as follows: single-peaked (s), double-peaked (d) and showing a P-Cygni profile (P), both regular or inverse. {An asterisk together with the variability indicator indicates that the source had been catalogued as UXOR type in the literature. Variability indicator values ($V_{i}$) for these objects could not be derived as they are not astrometrically well behaved. Similarly, many of these sources fall outside the Pre-Main Sequence tracks and isochrones in the HR diagram and no masses or ages could be derived for them. We decided to present the masses and ages in the cases they were computable but these values have to be taken with caution.} References for EW values and line shapes: (1) \citet{Fairlamb2}; (2) \citet{Carmona}; (3) \citet{Mendigutia3}; (4) \citet{Ababakr2}; (5) \citet{Hernandez}; (6) \citet{Baines}; (7) \citet{Wheelwright}; (8) \citet{van den Ancker3}; (9) \citet{Oudmaijer3}; (10) \citet{Kucerova}; (11) \citet{Hernandez2}; (12) \citet{Dunkin}; (13) \citet{Pogodin}; (14) \citet{Miroshnichenko3}; (15) \citet{Sartori}; (16) \citet{Polster}; (17) \citet{Manoj}; (18) \citet{Miroshnichenko4}; (19) \citet{Miroshnichenko}; (20) \citet{Borges}; (21) \citet{Vieira}; (22) \citet{Boehm}; (23) \citet{Nakano}; (24) \citet{Spezzi}; (25) \citet{Hou}; (26) \citet{Grinin2}; (27) \citet{Vieira2}; (28) \citet{Acke}; (29) \citet{Herbig2}; (30) \citet{Oudmaijer7}; (31) X-Shooter spectra, 2015, priv. comm, from ESO observing program 084.C-0952A; (32) \citet{Ababakr};
(33) \citet{Zuckerman}; (34) \citet{Oudmaijer7}; (35) \citet{Frasca}; (36) \citet{Miroshnichenko5}; (37) \citet{Miroshnichenko6}.}
\end{landscape}

\newpage

\renewcommand{\arraystretch}{1.2}
\begin{longtable}{lrrrrrrr}
\caption{\label{table4_2} IR excess at each bandpass (defined as F\textsubscript{observed}/F\textsubscript{CK}) for each Herbig Ae/Be star belonging to the low quality sample of 34 sources.}\\
\hline\hline 
Name & J &  H & K\textsubscript{s} & W1 & W2 & W3 & W4\\
  & $1.24\mu$m  & $1.66\mu$m & $2.16\mu$m & $3.4\mu$m & $4.6\mu$m & $12\mu$m & $22\mu$m\\ 
\hline
\endfirsthead
\caption{continued.}\\
\hline\hline
Name & J &  H & K\textsubscript{s} & W1 & W2 & W3 & W4\\
  & $1.24\mu$m  & $1.66\mu$m & $2.16\mu$m & $3.4\mu$m & $4.6\mu$m & $12\mu$m & $22\mu$m\\  
\hline
\endhead
\hline
\endfoot
BP Psc & $1.49$ & $2.28$ & $4.20$ & $10.16$ & $22.35$ & $168.78$ & $2200.40$ \\
DK Cha & $14.73$ & $49.13$ & $170.17$ & $325.14$ & - & $7497.91$ & $50750.83$ \\
GSC 5360-1033 & $2.65$ & $6.78$ & $12.87$ & - & - & - & - \\
GSC 5988-2257 & $2.62$ & $6.21$ & $16.62$ & $47.97$ & $117.47$ & $1245.27$ & $6594.26$ \\
GSC 6542-2339 & $2.19$ & $3.64$ & $5.68$ & - & - & - & - \\
HBC 1 & $88.74$ & $484.43$ & $2157.04$ & $9200.72$ & $27745.90$ & $156330.65$ & $1383272.17$ \\
HBC 324 & $0.93$ & $1.57$ & $3.60$ & $15.52$ & $54.31$ & $943.29$ & $13361.89$ \\
HBC 694 & $3.40$ & $8.51$ & $21.52$ & $103.52$ & $255.39$ & $1995.47$ & $31718.13$ \\
HBC 717 & $1.49$ & $2.35$ & $4.58$ & $12.21$ & $23.09$ & $149.10$ & $1086.81$ \\
HD 245906 & $1.65$ & $2.56$ & $4.15$ & $6.60$ & $9.67$ & $58.51$ & $465.93$ \\
HD 53367 & $1.27$ & $1.32$ & $1.39$ & $3.04$ & $4.84$ & $5.37$ & $116.70$ \\
HD 72106B & $1.79$ & $1.92$ & $2.73$ & $5.48$ & $10.88$ & $138.92$ & $976.86$ \\
Hen 2-80 & $2.60$ & $6.89$ & $17.96$ & $64.65$ & $211.29$ & $2910.69$ & $15743.40$ \\
MWC 314 & $2.32$ & $2.65$ & $3.51$ & $2.91$ & $6.01$ & $7.36$ & $23.11$ \\
MWC 623 & $4.04$ & $7.46$ & $12.17$ & $28.53$ & $78.51$ & $193.41$ & $397.67$ \\
MWC 930 & $0.35$ & $0.33$ & $0.35$ & $0.48$ & $0.86$ & $0.74$ & $2.16$ \\
NX Pup & $1.51$ & $4.32$ & $13.40$ & $50.79$ & $137.41$ & $359.52$ & $1521.35$ \\
PDS 144S & $3.81$ & $14.93$ & $58.24$ & $317.57$ & $681.57$ & $2883.63$ & $21943.52$ \\
PDS 229N & $1.36$ & - & - & $11.85$ & $22.39$ & $173.19$ & $938.57$ \\
PDS 322 & $1.08$ & $1.24$ & $1.58$ & $3.81$ & $8.96$ & $181.89$ & $22064.34$ \\
PDS 364 & $1.95$ & $4.35$ & $10.24$ & $25.65$ & $54.14$ & $737.90$ & $10364.41$ \\
PDS 371 & $1.03$ & $3.51$ & $9.47$ & $22.73$ & $55.17$ & $247.26$ & $2992.30$ \\
PDS 453 & $2.35$ & $3.96$ & $7.79$ & $22.88$ & $44.93$ & $279.44$ & $4685.85$ \\
PDS 530 & $0.86$ & $3.03$ & $13.43$ & $62.07$ & $146.37$ & $960.03$ & $7986.35$ \\
PDS 551 & $32.07$ & $111.81$ & $327.37$ & $1230.71$ & $4108.60$ & $29748.04$ & $220447.30$ \\
PDS 581 & $1.71$ & $9.18$ & $41.39$ & $207.15$ & $1047.80$ & $4498.40$ & $47015.82$ \\
PV Cep & $2.83$ & $24.69$ & $147.34$ & $483.75$ & $2383.40$ & $21362.00$ & $157219.18$ \\
R CrA & $18.56$ & $89.19$ & $567.83$ & - & - & $12353.87$ & $184275.03$ \\
UY Ori & $1.70$ & $3.77$ & $11.11$ & $36.23$ & $84.99$ & $1932.72$ & $12539.63$ \\
V590 Mon & $1.88$ & $4.42$ & $12.28$ & $44.17$ & $104.49$ & $2075.86$ & $21654.33$ \\
V645 Cyg & $0.74$ & $2.67$ & $19.54$ & - & - & $12523.05$ & $124226.58$ \\
V892 Tau & $18.38$ & $57.37$ & $142.10$ & $149.55$ & $392.45$ & $14054.94$ & $136211.22$ \\
VY Mon & $2.70$ & $5.67$ & $13.53$ & $38.47$ & - & $897.71$ & $7399.24$ \\
Z CMa & $1.21$ & $2.85$ & $9.21$ & - & - & $373.86$ & $2204.64$ \\
\end{longtable}

\end{document}